\documentclass[sigconf]{acmart}

\copyrightyear{2025}
\acmYear{2025}
\setcopyright{acmlicensed}
\acmDOI{XXXXXXX.XXXXXXX}
\acmConference[Conference acronym 'XX]{Make sure to enter the correct
  conference title from your rights confirmation email}{June 03--05,
  2018}{Woodstock, NY}
\acmISBN{978-1-4503-XXXX-X/2018/06}

\usepackage{amsmath}
\usepackage{setspace}
\usepackage{multirow}
\usepackage{tabularx}
\usepackage{color}
\usepackage{subfigure}
\usepackage{enumitem}
\usepackage{rotating}
\usepackage{bm}
\usepackage{float}
\usepackage[ruled,linesnumbered]{algorithm2e}

\usepackage{textcomp}
\usepackage{booktabs}
\usepackage{hyperref}
\DeclareMathOperator*{\argmax}{arg\,max}

\begin{document}

\title[CDC]{CDC: Causal Domain Clustering for Multi-Domain Recommendation}

\author{Huishi Luo}
\email{hsluo2000@buaa.edu.cn}
\orcid{0000-0002-3553-2280}
\affiliation{%
  \department{Institute of Artificial Intelligence}
  \institution{Beihang University}
  \city{Beijing}
  \country{China}
}

\author{Yiqing Wu}
\email{wuyiqing20s@ict.ac.cn}
\orcid{0000-0002-8068-9420}
\affiliation{%
  \institution{Institute of Computing Technology, Chinese Academy of Sciences}
  \city{Beijing}
  \country{China}
}

\author{Yiwen Chen}
\email{yiwenchen@buaa.edu.cn}
\orcid{0009-0008-0474-5209}
\affiliation{%
  \department{Institute of Artificial Intelligence}
  \institution{Beihang University}
  \city{Beijing}
  \country{China}
}

\author{Fuzhen Zhuang}
\authornote{Corresponding author.}
\email{zhuangfuzhen@buaa.edu.cn}
\orcid{0000-0001-9170-7009}
\affiliation{%
  \department{Institute of Artificial Intelligence}
  \institution{Beihang University}
  \city{Beijing}
  \country{China}
}

\author{Deqing Wang}
\email{dqwang@buaa.edu.cn}
\orcid{0000-0001-6441-4390}
\affiliation{%
  \department{SKLSDE, School of Computer Science}
  \institution{Beihang University}
  \city{Beijing}
  \country{China}
}
\renewcommand{\shortauthors}{Huishi Luo, Yiqing Wu, Yiwen Chen, Fuzhen Zhuang, and Deqing Wang}

\begin{abstract}
Multi-domain recommendation leverages domain-general knowledge to improve recommendations across several domains. However, as platforms expand to dozens or hundreds of scenarios, training all domains in a unified model leads to performance degradation due to significant inter-domain differences. Existing domain grouping methods, based on business logic or data similarities, often fail to capture the true transfer relationships required for optimal grouping. To effectively cluster domains, we propose Causal Domain Clustering (CDC). CDC models domain transfer patterns within a large number of domains using two distinct effects: the Isolated Domain Affinity Matrix for modeling non-interactive domain transfers, and the Hybrid Domain Affinity Matrix for considering dynamic domain synergy or interference under joint training. To integrate these two transfer effects, we introduce causal discovery to calculate a cohesion-based coefficient that adaptively balances their contributions. A Co-Optimized Dynamic Clustering algorithm iteratively optimizes target domain clustering and source domain selection for training. CDC significantly enhances performance across over 50 domains on public datasets and in industrial settings, achieving a 4.9\% increase in online eCPM. Code is available online\footnote{\href{https://github.com/Chrissie-Law/Causal-Domain-Clustering-for-Multi-Domain-Recommendation}{https://github.com/Chrissie-Law/Causal-Domain-Clustering-for-Multi-Domain-Recommendation}}.
\end{abstract}




\maketitle

\section{Introduction}
\label{sec:intro}

Personalized recommender systems aim to suggest potentially attractive items based on users' profiles and behavior histories, playing a crucial role in alleviating information overload and enhancing user experience. As platforms such as Amazon and Taobao expand, the number of recommendation domains has grown dramatically, now encompassing hundreds, from homepage and category-level recommendations to specialized marketing pages\cite{chang2023pepnet}. To effectively manage these domains, \textit{Multi-domain recommendation (MDR)} has been developed to capture domain-general knowledge and improve recommendation across various domains. Methods like STAR\cite{sheng2021star} and HiNet\cite{zhou2023hinet} adopt parameter-sharing strategies, explicitly separating domain-specific and domain-general knowledge to mitigate negative transfer\cite{zhuang2020survey}. Others, such as DTRN\cite{liu2023dtrn} and PEPNet\cite{chang2023pepnet}, use hypernetwork-inspired structures\cite{ha2016hypernetworks} to dynamically adapt network parameters based on discriminative domain features.

However, most existing MDR methods are designed for fewer than ten domains\cite{wang2022causalint, zhang2022sass, chang2023pepnet, jia2024d3}, and face substantial computational and memory burdens when scaled to dozens or hundreds of domains. Furthermore, these methods fundamentally assume domains are highly similar, asserting that joint domain training is more beneficial than training domains independently in terms of knowledge transfer. This assumption, however, often fails in industrial settings. For example, transferring knowledge between weakly related domains like electronics and cosmetics is challenging and can induce severe negative transfer. Joint training of these loosely related domains tends to degrade recommendation performance, a problem that substantially worsens as the number of domains increases.

\begin{figure}[!t]
\centering
\vspace{-8pt}
\includegraphics[width=1.0\linewidth,trim=0 12 0 20 ,clip]{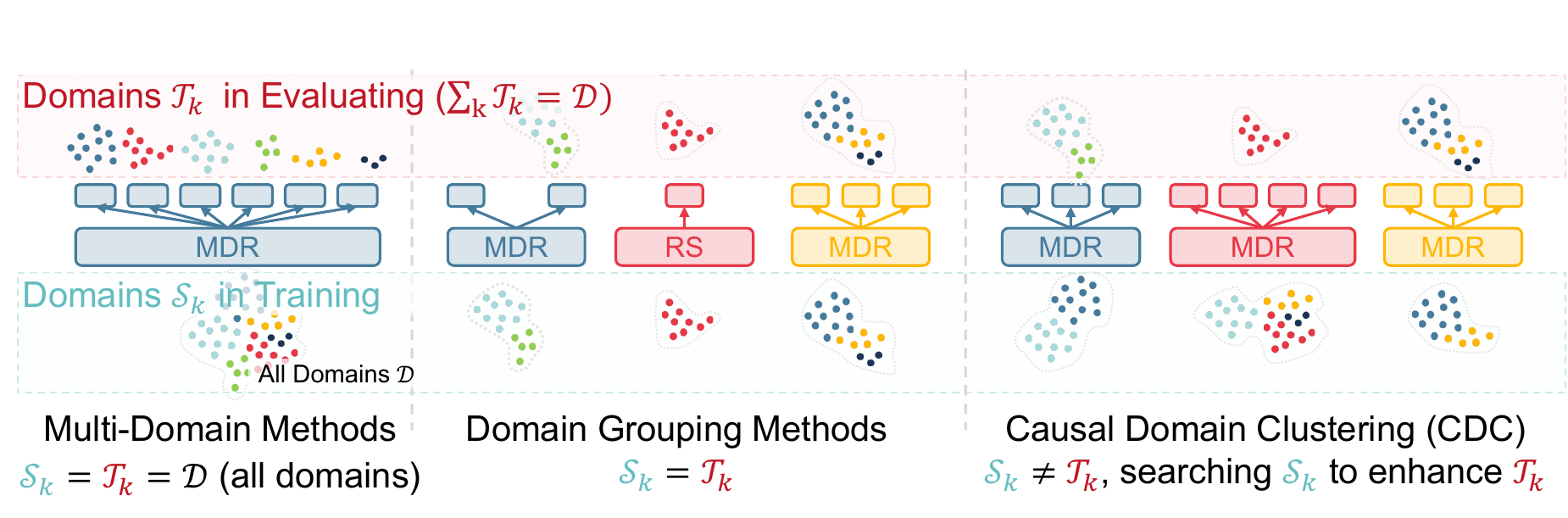} 
\vspace{-18pt}
\caption{Compared with existing MDR and domain grouping methods, CDC optimizes training source domain sets to maximize target cluster performance.}
\vspace{-16pt}
\label{fig:domain_clustering}
\end{figure}

To reduce computational complexity and enhance overall recommendation performance, domain clustering, which involves grouping a vast number of domains into several clusters, has emerged as a necessary and effective step before deploying MDR\cite{li2023adl}. Using the generated grouping strategy, developers can treat multiple domains within the same cluster as a single domain, thereby significantly reducing training resource overhead. Furthermore, it allows developers to train models independently within each group, without the distraction of irrelevant knowledge from other groups, thus enhancing both training efficiency and recommendation accuracy. Figure \ref{fig:domain_clustering} illustrates this process. Nevertheless, domain clustering for MDR encounters the following three primary challenges:

Challenge 1: \textbf{How to measure transfer relationships between domains?} Manually grouping domains for MDR often overlooks the inherent characteristics of data distributions. Although data distribution correlation~\cite{li2023adl, swayamdipta2020dataset, sherif2024stg} and gradient differences~\cite{bai2022saliency} are frequently used as an indicator of inter-domain relationships, they fail to adequately reflect the efficacy of knowledge transfer after clustering~\cite{standley2020tasks}. For example, a domain encompassing a wide variety of items may have a distribution that significantly differs from its more specialized sub-categories. However, joint training often provides these sub-category domains with valuable general knowledge from the broader domain, thereby improving recommendation performance. This observation underscores a critical discrepancy: data distributions are static reflections of domain characteristics, whereas the true effectiveness of domain knowledge transfer emerges only \textit{after training}. This often-overlooked gap arises from the predictive mapping capabilities of recommendation models. It leads to a misalignment between domain relevancy derived from data distributions or gradient similarities, and the true inter-domain relationships essential for optimal clustering. Therefore, it is essential to develop a more precise measure of one-to-one inter-domain transfer relationships that genuinely reflects the post-clustering performance.


Challenge 2: \textbf{How to evaluate transfer relationships considering domain interactions?} Measuring transfer relationships is further complicated by potential synergies or interferences among domains during higher-order joint training, collectively termed \textit{domain interaction}. For instance, a generalized apparel scenario might suffer negative transfer if trained only with men’s or women’s fashion due to distinct user characteristics. Training with both domains, however, could approximate the global distribution and foster synergistic interactions between the domains, yielding better results. It's crucial to consider the impact of domain synergy and interference in MDR; however, conventional one-to-one transfer metrics do not capture these interactions, or merely approximate them by combinations of lower-order relationships \cite{fifty2021efficiently}.


Challenge 3: \textbf{How to determine the optimal clustering and the best training data for each cluster?} Ideal clustering aims to maximize performance by jointly training domains within each cluster. However, given domain interaction and data scale impacts, it is nearly impossible to predict the final recommendation performance before completing the clustering, even with known one-to-one transfer effects between domains. Furthermore, due to the directional nature of domain transfers, the best training domain set for a cluster might exclude some of its own domains. Therefore, an effective domain clustering algorithm for MDR must not only identify the optimal target domain clustering but also select the corresponding source domains for training.


To address these challenges, we propose a novel \textbf{Causal Domain Clustering (CDC)} method for multi-domain recommendation. Specifically, by using one-step-forward (i.e., \textit{lookahead}) updates~\cite{fifty2021efficiently} with dynamic training domain sets, we explicitly model two domain transfer effect matrices. These matrices, based on loss function variations, bridge the traditional gap between optimization objectives and measurements based solely on data distribution (Challenge 1). To elaborate, the Isolated Domain Affinity Matrix captures transfer effects between domains under non-interaction conditions, while the Hybrid Domain Affinity Matrix considers the domain synergy or interference during joint training. To integrate these two transfer effects, we introduce causal discovery \cite{pearl2009causality} to calculate a cluster cohesion-based interaction coefficient, adaptively adjusting the two matrices' contribution to the final transfer effect (Challenge 2). Finally, our Co-Optimized Dynamic Clustering (CODC) algorithm heuristically solves optimal clusterings and corresponding source domain sets, providing a practical solution to an otherwise intractable problem (Challenge 3). In our experiments, CDC outperforms baselines on two public datasets and shows effectiveness in both offline and online settings on industrial platforms. The contributions of this work are summarized as follows:







(1) We propose a novel Causal Domain Clustering (CDC) framework, to the best of our knowledge, the first to incorporate causal discovery for domain clustering in multi-domain recommendation.

(2) We model the transfer effects between domains in a dual-dynamic manner. First, we dynamically adjust the training domains for updating the transfer matrices. Second, we propose a cohesion-based coefficient based on causal discovery, to dynamically evaluate the degree of domain synergy within the training domain set.

(3) We introduce the Co-Optimized Dynamic Clustering (CODC) algorithm, which simultaneously optimizes both the target domain clustering and the optimal source training domain set, enabling the best recommendation performance in the target domain.

(4) We achieve significant improvements on evaluations on two public and two industrial datasets. CDC is model-agnostic and easy to deploy. Over 14 days of A/B testing online across 64 domains, it resulted in a 4.9\% improvement in eCPM.

\section{Related Work}
\subsection{Multi-Domain Recommendation}

Multi-domain recommendation (MDR) uses domain knowledge transfer to enhance performance across all included domains. Inspired by multi-task learning \cite{ma2018mmoe, tang2020ple}, many methods, such as STAR \cite{sheng2021star}, AESM\textsuperscript{2} \cite{zou2022AESM2}, and HiNet \cite{zhou2023hinet}, explicitly partition shared and specific parameters to disentangle domain-specific and domain-general knowledge. Other approaches, including SAR-Net \cite{shen2021sarnet}, PEPNet \cite{chang2023pepnet}, and D3 \cite{jia2024d3}, focus on learning domain-discriminative feature representations with mechanisms like hyper-networks or attention. However, these studies typically require manually selecting domains in advance under the assumption that they are strongly relevant, an assumption that may not hold in real-world settings. In this paper, CDC addresses the challenge of clustering dozens or hundreds of domains, iteratively searching for optimal domain groups suitable for joint training. 
After clustering, any existing MDR method can be employed within each optimized cluster.

\subsection{Domain Clustering}

Domain clustering, or domain grouping, addresses the problem of determining which domains should be trained together in MDR. This issue has a significant impact on both the final performance and computational cost, yet remains relatively underexplored. A closely related research problem is task grouping in multi-task learning, where tasks are organized into groups to mitigate negative transfer. Many multi-task learning approaches adjust task loss weights to balance the multiple tasks~\cite{liu2019loss, raychaudhuri2022controllable, liu2022autolambda, liu2023multitask} and benefit from easy deployment. However, even with low task weights, training unrelated tasks together often leads to a seesaw effect~\cite{tang2020ple} and prevents achieving global optimality. Highly relevant are two-stage task grouping methods~\cite{standley2020tasks, fifty2021efficiently, wang2024principled}, which first learn task \textit{affinities} (i.e., task transfer relationships) during training and then employ search techniques like branch-and-bound~\cite{fifty2021efficiently} to find optimal groupings. In these studies, gains of higher-order task combinations are approximated based on pairwise relationships. In contrast, Song et al.~\cite{song2022efficient} utilized meta-learning to estimate task grouping gains. Our work extends Fifty et al.~\cite{fifty2021efficiently} by capturing domain affinities in both isolated and interaction settings, which addresses the challenge of evaluating high-order gains. We further introduce causal discovery to integrate these affinities. Beyond standard domain clustering, our method seeks to identify optimal training source domains that, though different from those within the target clusters, enhance the overall prediction performance of the grouped domains.

\section{Preliminary}

\subsection{Problem Definition}



We focus on the domain clustering phase in MDR, a critical step before model training. Given \(D\) domains \(\mathcal{D} = \{d_1, d_2, \ldots, d_D\}\), we aim to partition them into \(K\) clusters. Unlike conventional domain clustering methods, we allow distinct target/inference domains and training domains. As shown in Figure \ref{fig:domain_clustering}, our model optimizes two key variables: the \textbf{\emph{target domain cluster}} \(\mathcal{T}_k\) (with \(\sum_{k=1}^{K} \mathcal{T}_k = \mathcal{D}\)), which denotes the domains targeted for enhancement within \(K\) separate MDR models; and the \textbf{\emph{training source domain set}} \(\mathcal{S}_k\), which represents the domains used for training each model. In our paradigm, \(\mathcal{S}_k \neq \mathcal{T}_k\), and \(\mathcal{S}_k\) is specifically optimized to maximize performance of its corresponding target domain cluster \(\mathcal{T}_k\). 
Let $\Theta_{\mathcal{S}_k}$ denote the parameters of the separate model trained on $\mathcal{S}_k$:
\begin{footnotesize}
\begin{equation}
\Theta_{\mathcal{S}_k}=\underset{\Theta_{\mathcal{S}_k}}{\arg \min } \sum_{d \in \mathcal{T}_k} \mathcal{L}(f(x, d), y(x, d)),
\end{equation}
\end{footnotesize}
where $\mathcal{L}$ is the loss function, $f$ the recommender model for input features $x$ and domain indicator $d$, and $y(x, d)$ the binary label. Our goal is to find $\{\mathcal{T}_k,\mathcal{S}_k\}_{k=1}^K$ that minimizes
\begin{footnotesize}
\begin{equation}
    \mathcal{L}_{\text{total}} = \sum_{k = 1}^{K}{
    \sum_{d \in \mathcal{T}_k}{\mathcal{L}(f_{\Theta_{\mathcal{S}_k}}(x, d), y(x, d))}
    },
\end{equation}
\end{footnotesize}
where $f_{\Theta_{\mathcal{S}_k}}$ represents the model parameterized by $\Theta_{\mathcal{S}_k}$.

\subsection{Causal Discovery}
\label{sec:kernel}

Causal discovery, a key technique in causality research~\cite{pearl2009causality}, aims to learn causal structures and discover causal relationships from data under specific assumptions~\cite{luo2024ci4rs}. For example, given gene expression levels across different time points, causal discovery can investigate causal relationships among genes, determining how one gene may influence the expression of others~\cite{iyer1999transcriptional, dhillon2003diametrical, dhillon2004kernel, markham2022distance}. These genes, however, may not belong to a single unified causal structure, but rather multiple ones, each with unique relationship patterns, referred to as \textit{causal structural heterogeneity}. To address this, current studies use causal clustering to group samples (genes here) based on causal distance, identifying structurally homogeneous subsets, and then focus causal learning within each cluster~\cite{liu2015reverse, markham2022distance}. In this study, we use the \textit{dependence contribution kernel} \(\kappa\) proposed by Markham et al.\footnote{Open source implementation: \href{https://causal.dev/code/dep_con_kernel.py}{https://causal.dev/code/dep\_con\_kernel.py}} \cite{markham2022distance} to compute the causal distances between domains in MDR.

\section{Method}

\subsection{Overall Framework}

\begin{figure*}[!t]
    \centering
    \begin{minipage}[b]{1\linewidth}
    \centering
    \vspace{-13pt}
    \includegraphics[width=0.86\linewidth, trim=-10 0 -10 0, clip]{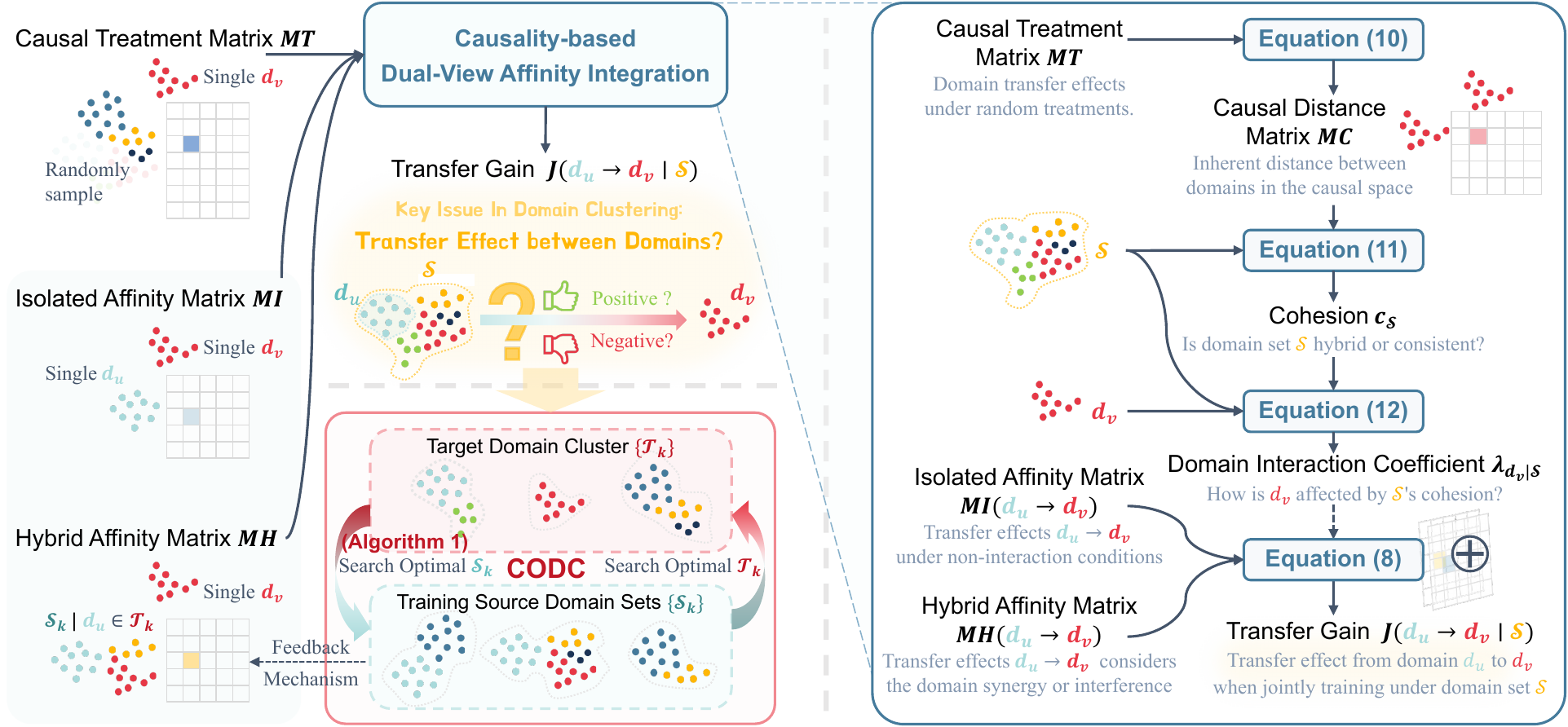}
    \end{minipage}
    \vspace{-18pt}
    \caption{Causal Domain Clustering (CDC) framework. (1) We learn two inter-domain affinity matrices, $MI$ and $MH$, to capture domain relationships independently and interactively. (2) Causal discovery computes an Interaction Coefficient \(\lambda\), enabling the adaptive integration of \(MI\) and \(MH\) to assess the transfer gain \(J\) under any possible training source domain set $\mathcal{S}$. (3) CODC jointly optimizes both the target domain clusters $\{\mathcal{T}_k\}$ and the optimal training source domain set $\{\mathcal{S}_k\}$.}
    \vspace{-16pt}
    \label{fig:cdc}
\end{figure*}

(1) As shown in Figure \ref{fig:cdc}, the Causal Domain Clustering (CDC) method first learns two distinct domain affinity matrices through pre-gradient updates. The Isolated Domain Affinity Matrix $MI$ quantifies the pairwise transfer effects across all domains without external influences, while the Hybrid Domain Affinity Matrix $MH$ considers the effects under domain interactions. (2) Next, causal discovery is employed to compute an Interaction Coefficient $\lambda$, which adaptively integrates the domain affinities from \(MI\) and \(MH\) to generate the transfer gain $J$. The transfer gain $J$ is then used to select the optimal training set $\mathcal{S}_k$ for the target domain clusters $\mathcal{T}_k$. (3) Finally, we propose the Co-Optimized Dynamic Clustering (CODC) algorithm, which alternately performs the following clustering steps: selecting the training source domain set $\mathcal{S}_k$ that maximizes performance for the target domain cluster $\mathcal{T}_k$; recalculating each domain's gain based on current training source domain sets to update $\mathcal{T}_k$. Note that CDC involves a feedback mechanism, allowing the chosen $\mathcal{S}_k$ to adapt the learning process of matrix $MH$. Therefore, as model training progresses, the reliability of the integrated gain \(J\) improves, leading to the convergence of the grouping strategy for both target domains and source domains.

\subsection{Inter-Domain Affinity Matrices}
\label{sec:matrix}
This section introduces two affinity matrices: the Isolated Domain Affinity Matrix $MI$ and the Hybrid Domain Affinity Matrix $MH$. Here, \textit{affinity} refers to the transfer effect calculated based on the loss difference. The matrices are designed to model domain relationships from two perspectives: independently for $MI$ and interactively for $MH$, providing a foundation for adaptive integration in Section \ref{sec:integrate}.
\subsubsection{Isolated Domain Affinity Matrix}
\label{sec:mi}
Effective domain clustering requires a metric to assess inter-domain relationships. Traditional grouping methods presume that domains with similar distributions should be clustered together, and often employ metrics such as static data distribution similarity \cite{swayamdipta2020dataset, sherif2024stg} or dynamic gradient distribution similarity during training \cite{bai2022saliency} to assess inter-domain affinity. However, as previously discussed in Challenge 1 of Section \ref{sec:intro} and corroborated by Standley et al. \cite{standley2020tasks}, similarity-based metrics may not be reliably effective for MDR scenarios because similarity does not necessarily equate to effective knowledge transfer. Moreover, these traditional metrics are non-directional (i.e., the influence of domain $d_u$ on domain $d_v$ is considered equivalent to that of domain $d_v$ on domain $d_u$), which fails to reflect the directional nature of knowledge transfer that is inherent in transfer learning.

Therefore, inspired by ~\cite{fifty2021efficiently,wang2024principled}, we employ loss changes during one-step-forward training (also known as the \textit{lookahead} method) to measure inter-domain affinity. The lookahead method leverages future information to guide the current state~\cite{wang2024principled}, involving three steps: saving the current state of the MDR base model, applying one or more steps of gradient update with data from a subset of domains, and then reloading the saved model parameters to compare performance differences~\cite{fifty2021efficiently}. Specifically, we first warm up the base model using several batches of training data. At a given time step $t$, we save the current model parameters $\Theta^t$. We then use data exclusively from domain $d_u$ to train and update the model's parameters, achieving $\Theta^{t+1}_{\{d_u\}}$. This updated model is subsequently employed to predict the loss for domain $d_v$. Let matrix $MI^t \in \mathbb{R}^{D \times D}$ represent overall isolated inter-domain affinity, where $MI^t[u, v]$ denotes the affinity of domain $d_u$ on domain $d_v$, denoted as:
\begin{footnotesize}
\begin{align}
\label{eq:mi}
    MI^{t}[u, v] = \alpha &MI^{t-1}[u, v] \nonumber \\
    \quad + (1-\alpha)&\left(1 - \frac{\sum\limits_{d\in d_v}{\mathcal{L}(f_{\Theta^{t+1}_{\{d_u\}}}(x, d), y(x, d))}}{\sum\limits_{d\in d_v}{\mathcal{L}(f_{\Theta^{t}}(x, d), y(x, d))}}\right),
\end{align}
\end{footnotesize}
where $\alpha$ is a smoothing factor to stabilize history affinity. 

A positive value of $MI^t[u, v]$ indicates that the update on the model parameters results in a lower loss on domain $d_v$ than the original parameter values, suggesting a beneficial transfer of knowledge from domain $d_u$ to $d_v$. Conversely, a negative value implies that the parameter update is adversely affects domain $d_v$’s performance, indicating a negative transfer effect. 


\subsubsection{Hybrid Domain Affinity Matrix}
\label{sec:mh}

For any two domains $d_u$ and $d_v$ in MDR, the transfer of beneficial knowledge from $d_u$ to $d_v$ can manifest in two ways. Firstly, there may be a direct, positive independent transfer gain, as indicated by a positive value in the Isolated Domain Affinity Matrix $MI[u, v]$. Secondly, a synergistic effect may arise when $d_u$ interacts with other domains during joint training, even if $MI[u, v]$ is not positive.

To quantify the interaction transfer effects, we introduce the Hybrid Domain Affinity Matrix $MH \in \mathbb{R}^{D \times D}$, designed to assess inter-domain affinity in a domain-interaction training context. At time step $t$, let $\mathcal{M}^t_u$ be the mixed domain set containing $d_u$. We perform a lookahead update of the model using data from domains in $\mathcal{M}^t_u$ and $\mathcal{M}^t_u - \{d_u\}$, obtaining parameters $\Theta^{t+1}_{\mathcal{M}^t_u}$ and $\Theta^{t+1}_{\mathcal{M}^t_u-\{d_u\}}$, respectively. Here, $\mathcal{M}^t_u - \{d_u\}$ denotes the set of domains in $\mathcal{M}^t_u$ excluding domain $d_u$. This design allows $\Theta^{t+1}_{\mathcal{M}^t_u}$ to capture all aspects of knowledge transfer from $d_u$, particularly its interactions with other domains, while $\Theta^{t+1}_{\mathcal{M}^t_u-\{d_u\}}$ excludes these influences. Consequently, $MH^t[u, v]$ is computed as:
\begin{footnotesize}
\begin{align}
\label{eq:mh}
    MH^{t}[u, v] = \alpha&MH^{t-1}[u, v] \nonumber \\
    \quad + (1-\alpha)&\left(1 - \frac{\sum\limits_{d\in d_v}{\mathcal{L}(f_{\Theta^{t+1}_{\mathcal{M}^t_u}}(x, d), y(x, d))}}{\sum\limits_{d\in d_v}{\mathcal{L}(f_{\Theta^{t+1}_{\mathcal{M}^t_u-\{d_u\}}}(x, d), y(x, d))}}\right),
\end{align}
\end{footnotesize}
where $\alpha$ serves as the smoothing factor, consistent with Equation (\ref{eq:mi}). The metric $MH^t[u, v]$ cleverly assesses the variation in $d_v$'s loss when the mixed domain set $\mathcal{M}^t_u$ either includes or excludes $d_u$, thus quantifying the affinity of $d_u$ towards $d_v$ under interactions. A positive $MH^t[u, v]$ indicates that including $d_u$ in the mixed domain set leads to parameter updates that benefit $d_v$, suggesting a synergistic effect between $d_u$ and other domains within the set. 

In practical training, the initial mixed domain set for domain $d_u$, $\mathcal{M}^t_u$, starts as the complete set $\mathcal{M}^0_u = \mathcal{D}$, and adapts over time based on the domain clustering outcomes at each time step $t$:
\begin{footnotesize}
\begin{equation}
\label{eq:U}
\mathcal{M}^t_u = \begin{cases} 
\mathcal{D}, & \text{if } t = 0, \\
\mathcal{S}^t_k, & \text{if } t > 0 \text{ and } k \text{ is such that } d_u \in \mathcal{T}^t_k.
\end{cases}
\end{equation}
\end{footnotesize}

The innovation of the Hybrid Domain Affinity Matrix lies in its strategy of excluding a specific training domain and then performing a lookahead update. This strategy enables the explicit demonstration of inter-domain interactions, illustrating how one domain either synergizes with or impedes other domains within a mixed domain set.
Furthermore, the mixed domain set $\mathcal{M}^t_u$ dynamically evolves through successive iterations of domain clustering (Equation (\ref{eq:U})). This \textit{feedback mechanism} enables $MH^t$ to progressively approach the true inter-domain affinity within the final clustering scheme, thereby mitigating the estimation bias of traditional clustering methods, where higher-order affinities are simply estimated by combinations of lower-order ones\cite{liu2022autolambda}.

\subsection{Causality-based Dual-View Affinity Integration}
\label{sec:integrate}

We propose a dual-view affinity integration method to fuse the affinities from $MI$ and $MH$ using causal discovery. To quantify intra-cluster similarity, we define \textit{cohesion} as the average causal distance between domains in a cluster. The method operates under two hypotheses: (1) When the cohesion among a jointly trained domain set is high, indicating strong domain similarity, beneficial knowledge transfer occurs mainly through direct, internal interactions. Thus, the Isolated Domain Affinity $MI$, which reflects affinity without inter-domain interactions, should be preferred. (2) Conversely, if cohesion is low, implying significant disparities among the domains, interactions across these domains-whether synergy or interference-can critically influence the transfer effectiveness. In such cases, reliance on the Hybrid Domain Affinity $MH$ is advised, as it accounts for the complex dynamics of domain interactions.

Based on these hypotheses, at time step $t$, given any possible training source domain set $\mathcal{S}$, our integration method follows three steps. First, we estimate the cohesion $c_{\mathcal{S}}$ of $\mathcal{S}$. Next, based on this cohesion, we evaluate its influence on the target domain $d_v$, quantified by the \textit{domain interaction coefficient} $\lambda_{d_v\mid \mathcal{S}}$, as the effect of $\mathcal{S}$ differs for different target domains. Lastly, using $\lambda_{d_v\mid \mathcal{S}}$, we compute the final transfer gain $J(u \rightarrow v \mid \mathcal{S})$ of domain $d_u$ on $d_v$:

\vspace{-2pt}
\begin{footnotesize}
\begin{equation}
\label{eq:J}
    J(d_u\rightarrow d_v \mid \mathcal{S})=(1-\lambda_{d_v\mid \mathcal{S}}) MI[u, v]+\lambda_{d_v\mid \mathcal{S}}MH[u, v].
\end{equation}
\end{footnotesize}

The domain interaction coefficient $\lambda_{d_v\mid \mathcal{S}}$, inversely proportional to \( \mathcal{S} \)'s cohesion, quantifies the influence of complexity within the training domain set $\mathcal{S}$ on domain $d_v$. This coefficient is crucial not only for calculating the transfer gain $J(d_u \rightarrow d_v \mid \mathcal{S})$ but for initializing clusters during the iterative clustering process (Equation (\ref{eq:init_S}) and Algorithm \ref{al:cdc}, Line 9). In the following sections, we detail how causal discovery theory is applied to calculate $\lambda_{d_v \mid \mathcal{S}}$, with random domain sampling used as \textit{Treatments} \cite{pearl2009causality} to derive the treatment effect matrix $MT$, and a causal distance matrix $MC$ computed to capture latent domain dependencies more effectively.

\subsubsection{Domain Causal Distance}
\label{sec:mc}

To compute the interaction coefficient \(\lambda\), a reliable domain distance metric is essential. Traditional metrics, such as distribution divergence, Euclidean distance, or cosine similarity, fail to capture dependencies in the transfer effect space. Instead, we adopt causal discovery, leveraging \textit{treatment effects} after \textit{intervention} to reveal underlying causal structures. For example, in genomics, causal discovery is utilized to determine causal distances between genes based on their expression levels, with genes considered strongly linked if they consistently co-express or co-suppress. Similarly, in multi-domain learning, two domains are considered similar if their performance gains remain consistently positive or negative across various training domains.

We approach the random sampling of domain data for training as a \textit{treatment} (or \textit{action}) in causality theory. Variations in domain losses are then interpreted as the \textit{treatment effect}. At time step $t$, we perform $R$ sampling actions, and the treatment effects for $D$ domains across these actions are recorded in a Causal Treatment Matrix $MT^t \in \mathbb{R}^{R \times D}$. For the $r^{th}$ random sampling, the corresponding domain set is denoted as $\mathcal{R}^t_r$, and the gain for domain $d_v$ in this set is represented by $MT^t[r, v]$:

\begin{footnotesize}
\begin{equation}
\label{eq:mt}
    MT^t[r, v] = 1 - \frac{\sum\limits_{d\in d_v}{\mathcal{L}(f_{\Theta^{t+1}_{\mathcal{R}^t_r }}(x, d), y(x, d))}}{\sum\limits_{d\in d_v}{\mathcal{L}(f_{\Theta^{t}}(x, d), y(x, d))}}.
\end{equation}
\end{footnotesize}
Following, we use the \textit{dependence contribution kernel} $\kappa$, introduced in Section \ref{sec:kernel}, to derive a symmetric Causal Distance Matrix $MC^t \in [0, \pi/2]^{D \times D}$ from the matrix $MT^t \in \mathbb{R}^{R \times D}$:
\begin{footnotesize}
\begin{equation}
\label{eq:dis}
    MC^t[u,v] = \arccos\left(\kappa(MT^t[:, u], MT^t[:, v])\right),
\end{equation}
\end{footnotesize}
where $MT^t[:, u]$ and $MT^t[:, v]$ denote the treatment effect vectors for domains $d_u$ and $d_v$ across various treatments, respectively. 

The kernel $\kappa$ is essentially a cosine similarity in the causal space; thus, $\arccos \circ \kappa$ quantifies the dissimilarity in causal dependence patterns between $d_u$ and $d_v$. The strength of this kernel lies in its isometry with the space of causal ancestral graphs \cite{markham2022distance}. Therefore, unlike traditional metrics such as cosine distance based on $MT^t[r, j]$, causal distance more accurately captures the inherent similarity between domains, as validated by experiments (see Section \ref{sec:ablation}). It is important to note that while domain clustering bears similarities with causal clustering, there are no actual causal relationships among domains. In other words, their interactions are not governed by stable physical mechanisms in nature.

\subsubsection{Cohesion and Interaction Coefficient}

Based on the causal distance $MC^t[u,v]$, we can calculate the cohesion of domain set $c^t_{\mathcal{S}}$ and domain interaction coefficient $\lambda_{d_v\mid \mathcal{S}}$. Given a time step $t$, the cohesion $c^t_{\mathcal{S}}$ of any given domain set $\mathcal{S}$ quantifies the average separation among domains within the set, as calculated from the causal distance $MC^t$:
\begin{footnotesize}
\begin{equation}
    c^t_{\mathcal{S}} = \frac{1}{\left|\mathcal{S}\right|^2}\sum_{d_u\in \mathcal{S}} \sum_{d_v\in \mathcal{S}} MC^t[u,v].
\end{equation}
\end{footnotesize}
A higher value of $c^t_{\mathcal{S}}$ indicates a greater causal distance among the domains, suggesting significant diversity within the set. Conversely, a lower value indicates a higher degree of similarity among the domains, implying a more cohesive set.

Based on this, the domain interaction coefficient $\lambda_{d_v\mid \mathcal{S}} \in [0, 1]$ for domain $d_v \in \mathcal{S}$, first introduced in Equation (\ref{eq:J}), is defined as:
\begin{footnotesize}
\begin{align}
\label{eq:lambda}
\lambda_{d_v\mid \mathcal{S}} & = \min\left(1, \frac{\sum_{d_u\in\mathcal{S}}MC^t[u,v]}{2\left|\mathcal{S}\right|c^t_{\mathcal{S}}}\right)  \nonumber \\ 
& = \min\left(1, \frac{\left|\mathcal{S}\right|\sum_{d_u\in\mathcal{S}}MC^t[u,v]}{2\sum_{d_u\in \mathcal{S}} \sum_{d_v\in \mathcal{S}} MC^t[u,v]}\right).
\end{align}
\end{footnotesize}
The interaction coefficient $\lambda_{d_v\mid \mathcal{S}}$ leverages relative distances to dynamically consider both the cohesion of the domain set $\mathcal{S}$ and its influence on the target domain $d_v$. According to Equations (\ref{eq:J}) and (\ref{eq:lambda}), when $\lambda_{d_v\mid \mathcal{S}}$ approaches $1$, it indicates that domain $d_v$ is significantly different from other domains in $\mathcal{S}$, making the Hybrid Domain Affinity Matrix $MH$ predominantly determine the final effect estimation. Conversely, when $\lambda_{d_v\mid \mathcal{S}}$ is close to $0$, $d_v$ closely aligns with the general domains in $\mathcal{S}$, emphasizing the Isolated Domain Affinity Matrix $MI$ in determining the transfer effect.

\subsection{Co-Optimized Dynamic Clustering}
\label{sec:codc}

Given the NP-hard nature of the domain clustering problem, we propose the heuristic Co-Optimized Dynamic Clustering (CODC) method. CODC iteratively solves both the domain groupings $\mathcal{T}_k$ and their corresponding optimal training domain sets $\mathcal{S}_k$.

\subsubsection{DM Training Domain Optimization}
\label{sec:dm}

The Dual-Metric (DM) Training Domain Optimization algorithm is designed to identify the optimal training domain groups $\{\mathcal{S}^t_k\}_{k=1}^{K}$ for predefined target domain groups $\{\mathcal{T}^t_k\}_{k=1}^{K}$, formalized by the function $F$:
\begin{footnotesize}
\begin{equation}
    \mathcal{S}^t_k=F(\mathcal{T}^t_k).
\end{equation}
\end{footnotesize}
As detailed in Algorithm \ref{al:search_train}, it combines two metrics: the inter-domain affinity $J(u \rightarrow v \mid \mathcal{S})$ and an additional \textit{Affiliation Score} $P$. $J$ measures the current transfer gain, while $P$ accounts for global benefits. This dual-metric strategy balances immediate performance improvements with long-term strategic fit. 

\SetAlgoNoEnd
\begin{algorithm}[!t]
\footnotesize
\setstretch{0.8}
\SetKwInOut{KwInit}{Init}
\caption{Dual-Metric Training Domain Optimization}
\label{al:search_train}
\KwIn{All domains $\mathcal{D} = \{d_1, \dots, d_D\}$; Domain sample proportions $\{w_u\}_{u=1}^D$, where $\sum_{u=1}^D w_u = 1$; Target domain cluster $\mathcal{T}_k^t$ for $k \in \{1, \dots, K\}$ at time step $t$; Domain affinity matrices $MI^t$ and $MH^t$; Causal distance matrix $MC^t$; Domain count $N$ to initialize set $\mathcal{S}^t_{k,0}$; Dynamic weight $\rho^t$ of affiliation score.
}
\KwOut{Optimal training source domain set $\mathcal{S}_k^t$ maximizing overall performance for $\mathcal{T}_k^t$.}
\KwInit{Set iteration counter $i \gets 0$; Initialize $\mathcal{S}^t_{k,0}$ using Equation (\ref{eq:init_S}).}
\While{$|\mathcal{S}^t_{k,i}| < D$}{
    \ForEach{candidate domain $d_u \notin \mathcal{S}^t_{k,i}$}{
        Calculate transfer gain $J(d_u \rightarrow \mathcal{T}_k^t | \mathcal{S}^t_{k,i} \cup \{d_u\})$ based on Equation (\ref{eq:J_T}), using $MI^t, MH^t, MC^t$;\\
        \eIf{$t = 0$}{
            $P(d_u | k) \gets 0$;
        }{
            Calculate affiliation score $P(d_u | k)$ based on Equation (\ref{eq:P}), using $MC^t, \mathcal{S}_k^0, w_u$;
        }
        Select the optimal candidate domain $d_u$ based on Equation (\ref{eq:JP}), using $\rho^t$ and the calculated $J$ and $P$; \\
        \If{$J(d_u \rightarrow \mathcal{T}_k^t | \mathcal{S}^t_{k,i} \cup \{d_u\}) + \rho^t \cdot P(d_u | k) < 0$}{
            break;
        }
        $\mathcal{S}^t_{k,i+1} \gets \mathcal{S}^t_{k,i} \cup \{d_u\}$; \\
        $i \gets i + 1$;
    }
}
$\mathcal{S}^t_{k} \gets \mathcal{S}^t_{k,i}$
\end{algorithm}
\normalsize

Initially, the training domain sets $\mathcal{S}^t_k$ are initialized by selecting the $N$ domains from $\mathcal{T}^t_k$ with the smallest $\lambda$ values:
\begin{footnotesize}
\begin{equation}
\label{eq:init_S}
    \mathcal{S}^t_{k,0} = \{ d_v \in \mathcal{T}^t_k \mid \lambda_{d_v \mid \mathcal{T}^t_k} \text{ is among the smallest } N \text{ values}\}.
\end{equation}
\end{footnotesize}
Subsequently, during each iteration $i$, we evaluate the transfer gain of each external domain $d_u \notin \mathcal{S}_{k,i}$ to the target domain set $\mathcal{T}_k$:
\vspace{-3pt}
\begin{footnotesize}
\begin{equation}
\label{eq:J_T}
    J(d_u\rightarrow \mathcal{T}^t_k \mid \mathcal{S}^t_{k,i}\cup\{d_u\} ) = \frac{\sum\limits_{d_v \in \mathcal{T}^t_k } w_v J(d_u\rightarrow d_v \mid \mathcal{S}^t_{k,i}\cup\{d_u\})}{\sum\limits_{d_v \in \mathcal{T}^t_k} w_v}  .
\end{equation}
\end{footnotesize}
Here, $w_v$ represents the proportion of samples in domain $d_v$, accounting for domain sample size to ensure fairness.

To enhance the robustness of $\mathcal{S}^t_{k,i+1}$ and prevent local optima that could impede subsequent optimization step (Equation (\ref{eq:U})), we introduce the domain \textit{Affiliation Score} $P(d_u \mid k) \in [-1,1]$. This score quantifies the likelihood of domain $d_u$ belonging to the final training domain set $\mathcal{S}_k$:

\begin{footnotesize}
\begin{equation}
\label{eq:P}
    P(d_u \mid k) = (1-2*\lambda_{d_u \mid \mathcal{S}_k^0})*\sqrt{w_u},
\end{equation}
\end{footnotesize}
where $\mathcal{S}^0_k$, derived from causal distance clustering, provides a reliable and robust reliable starting point for further optimizations. 

\SetAlgoNoEnd
\begin{algorithm}[!t]
\footnotesize
\setstretch{0.8}
\SetKwInOut{KwInit}{Init}
\caption{Causal Domain Clustering (CDC)}
\label{al:cdc}
\KwIn{Base multi-domain model of CDC with parameter $\Theta$; All domains $\mathcal{D} = \{d_1, \dots, d_D\}$.}
\KwOut{Target domain clusters $\{\mathcal{T}_k\}_{k=1}^K$ and training source domain sets $\{\mathcal{S}_k\}_{k=1}^K$; Optimized CDC model.}
\KwInit{Warm up CDC; set time step $t \gets 0$.}

\While{CDC has not converged}{
  \eIf{condition to update domain matrices is met}{
    Compute $MI^t$, $MH^t$, $MT^t$ based on the lookahead method\;
    Calculate matrix $MC^t$ using $MT^t$ based on Equation (\ref{eq:dis})\;

    \eIf{$t = 0$}{
        Perform $K$-means clustering on $MC^t$ to get $\{\mathcal{T}^t_k\}_{k=1}^K$\;
    }{
        \For{$k \gets 1$ \KwTo $K$}{
            Initialize $\mathcal{T}^t_k \gets \{d_v\in \mathcal{T}^{t-1}_k\}$ where $\lambda_{d_v \mid \mathcal{T}^{t-1}_k}$ is the smallest;
            }
        Set unclustered domain set $\mathcal{Q} \gets \mathcal{D} - \bigcup_{k=1}^K \mathcal{T}^t_k$\;
        \While{$\mathcal{Q}$ is not empty}{
            \For{$k \gets 1$ \KwTo $K$}{
                Compute $\mathcal{S}^t_k \gets F(\mathcal{T}^t_k)$ using Algorithm \ref{al:search_train}\;
                Find candidate domain for $\mathcal{T}^t_k$, i.e., $d_{cand,k} = \underset{d_v \in \mathcal{Q}}{\arg \max } J(\mathcal{S}^t_k \rightarrow d_v)$, where $J(\mathcal{S}^t_k \rightarrow d_v) = \sum_{d_u \in \mathcal{S}^t_k} J(d_u \rightarrow d_v \mid \mathcal{S}^t_k)$\;
            }
            \For{$k \gets 1$ \KwTo $K$}{
                \tcp{\scriptsize Verify if \(k\) is the most optimal cluster for domain \(d_{cand,k}\)}
                \If{$k = \underset{k=1, \dots, K}{\arg \max } J(\mathcal{S}^t_k \rightarrow d_{cand,k})$}{
                    $\mathcal{T}^t_k \gets \mathcal{T}^t_k \cup \{d_{cand,k}\}$\;
                    $\mathcal{Q} \gets \mathcal{Q} - \{d_{cand,k}\}$\;
                }
            }
        }
    }
    Compute $\{\mathcal{S}^t_k \gets F(\mathcal{T}^t_k)\}_{k=1}^K$ using Algorithm \ref{al:search_train}\;
    $t \gets t + 1$\;
  }{Train the CDC model according to the domain cluster \(\{\mathcal{T}^t_k\}_{k=1}^{K}\). Specifically, for sample from domain \(d\), update the shared parameters and the \(k\)-th domain-specific parameters, where \(k\) is determined by the cluster \(d \in \mathcal{T}^t_k\)\;
  }
}
\end{algorithm}
\normalsize
Finally, the training domain set for the next iteration is updated by combining Equations (\ref{eq:J_T}) and (\ref{eq:P}):
\begin{footnotesize}
\begin{align}
    d_u=\argmax_{d_u\notin \mathcal{S}^t_{k,i} } & \left(J(d_u\rightarrow \mathcal{T}^t_k \mid \mathcal{S}^t_{k,i}\cup\{d_u\} )+\rho^t   P(d_u \mid k) \right), \label{eq:JP} \\
    & \mathcal{S}_{k, i+1}^t = \mathcal{S}_{k, i}^t \cup \{d_u\},
\end{align}
\end{footnotesize}
where the decay coefficient $\rho^t = \rho^0 * \beta^t $  gradually reduces the influence of $P$ over time to prioritize transfer gains.


Algorithm \ref{al:search_train} outlines the complete DM Training Optimization process. It dynamically optimizes training domain selection by simultaneously balancing the immediate transfer gains and global affiliation scores. This approach mitigates the volatility of early optimization gains and enhances the robustness.

\subsubsection{Heuristic Co-Optimization}
\label{sec:heuristic}

As outlined in Algorithm \ref{al:cdc}, the heuristic Co-Optimized Dynamic Clustering (CODC) algorithm iteratively executes the following clustering steps: selecting the optimal source training domain set $\mathcal{S}^t_k$ for maximum performance of target domain group $\mathcal{T}^t_k$; and based on the current groupings of source training domains $\mathcal{S}^t_k$, computing the gains for each domain to assign them to the target group where they maximize the gain.

\section{Experiments}
\label{sec:exp}

\subsection{Experimental Setup}

\textbf{Public Dataset.} We use two widely recognized datasets: the Amazon dataset \cite{ni2019justifying} and AliCCP dataset \cite{ma2018entire}, where domains are defined based on item categories. Detailed statistics of these datasets are provided in Table \ref{tab:dataset}. The \textbf{Amazon dataset} includes all $25$ item categories, with data from the most recent $12$ months. Ratings above $4$ are labeled as positive. The data is temporally split into training, validation, and test sets with a ratio of $90:5:5$. For the \textbf{AliCCP dataset}, we sample $50$ item categories as distinct domains. To more closely simulate real-world recommender platforms, where a domain typically contains multiple item categories, an additional $10$ item categories are randomly selected and merged with existing domains. Click events serve as binary labels in this dataset. The training set uses the dataset's predefined training segment, while the original test set is split equally into validation and test subsets.

\begin{table}[!t]
  \centering
  \renewcommand{\arraystretch}{0.75}
  \vspace{-9pt}
  \caption{Public datasets statistics. The "Majority ratio" refers to the sample ratio in the largest domain, and "Minor domains" represent domains with less than 2\% of the samples.}
  \vspace{-8pt}  
  \setlength{\tabcolsep}{3.3mm}
  {
    \footnotesize
    \begin{tabular}{crr}
    \toprule
    Datasets & Amazon & AliCCP \\
    \midrule
    \#Users & 2,899,335  & 234,952  \\
    \#Items & 1,286,361  & 107,694  \\
    \#Samples & 17,664,862  & 18,952,318  \\
    \#Positive samples & 12,013,077  & 834,050  \\
    \#Domains ($D$) & 25    & 50 \\
    Majority ratio & 16.99\%  & 14.26\% \\
    \#Minor domains & 12    & 38 \\
    \bottomrule
    \end{tabular}}%
  \label{tab:dataset}%
  \vspace{-16pt}
\end{table}%

\textbf{Industrial Dataset.} For offline evaluations, we collect exposure logs of \textbf{81 business domains} from the recommender system of an industrial app marketplace, which \textbf{ranks among the top five globally by monthly active users}. These domains feature relatively high user interaction costs and low data volumes. To enhance their performance, we incorporate auxiliary data from three high-volume domains. Specifically, we collect datasets over two distinct periods: MDR-229M spanning $8$ days with $229$ million records, and MDR-865M over $31$ days with $865$ million records. Download logs are defined as positive samples and make up $4.73\%$ of the data. The three high-volume domains account for $63.67\%$ of the total volume, while 50 domains each contribute less than $0.5\%$. The final day’s data is reserved for testing, with the remainder used for training.

\subsection{Baselines}
\label{sec:baseline}
We compare CDC with a range of state-of-the-art baselines, categorized into three types based on their transfer learning strategies.

(1) \textbf{Single-Domain Recommendation Models}: \textbf{DeepFM}~\cite{guo2017deepfm}, \textbf{DCN}~\cite{wang2017dcn}, \textbf{AutoInt}~\cite{song2019autoint} and \textbf{DCNv2}~\cite{wang2021dcnv2}, each trained on the mixed dataset aggregated from all domains.

(2) \textbf{Multi-Domain Recommendation Models}: We adapt multi-task recommenders such as \textbf{MMoE}~\cite{ma2018mmoe} and \textbf{PLE}~\cite{tang2020ple} by assigning a domain-specific tower network to each domain. We also consider MDR-specific methods: \textbf{STAR}~\cite{sheng2021star}, \textbf{AdaSparse}~\cite{yang2022adasparse}, \textbf{HiNet}~\cite{zhou2023hinet}, \textbf{PEPNet}~\cite{chang2023pepnet}, and \textbf{ADL}~\cite{li2023adl}.

(3) \textbf{Domain Grouping Methods}: These methods, including \textbf{Random}, \textbf{Manual} (not applicable to the AliCCP dataset due to anonymized domains), and \textbf{TAG}~\cite{fifty2021efficiently}, implement MDR in two phases: first, domains are explicitly grouped into $K$ clusters based on specific criteria or strategies; second, the $K$ clusters are treated as $K$ individual domains for multi-domain learning. 
Our CDC method supports both a similar two-phase learning process (\textbf{CDC}) and a streamlined, end-to-end framework (\textbf{CDC (end-to-end)}). Notably, some recent grouping-related methods, such as MTG-Net~\cite{song2022efficient}, are excluded due to their high computational overhead (requiring over 3000 meta‑training cycles). Instead, we compare CDC with the SOTA baseline TAG, which sufficiently demonstrates CDC’s advancements, as supported by recent literature \cite{wang2024principled}.

\begin{table*}[t]
  \centering
  \renewcommand{\arraystretch}{0.9}
  \vspace{-14pt}
  \caption{Performance comparison of different methods trained across all domains on Amazon and AliCCP datasets. Best and second-best results are highlighted in bold and underlined, respectively. $*$ indicates statistically significant differences (\textit{p}-value $< 0.01$) from the second-best baseline. Results are averaged over five runs.}
  \vspace{-8pt}
  \setlength{\tabcolsep}{6pt}
  {
    \footnotesize
    \begin{tabular}{clcccccccc}
    \toprule
    \multicolumn{2}{c}{\multirow{2}[4]{*}{Method}} & \multicolumn{4}{c}{Amazon}    & \multicolumn{4}{c}{AliCCP} \\
    \cmidrule(lr){3-6} \cmidrule(lr){7-10} \multicolumn{2}{c}{}  & DomainAUC & TotalAUC   & Major5AUC  & \multicolumn{1}{c}{Minor10AUC } & DomainAUC & TotalAUC   & Major5AUC  & Minor10AUC  \\
    \midrule
    \multicolumn{1}{c}{\multirow{4}[2]{*}{\begin{sideways}Single\end{sideways}}} & DeepFM & 0.6451  & 0.6394  & 0.6490  & 0.6515  & 0.5764  & 0.5908  & 0.5768  & 0.5645  \\
          & DCN   & 0.6524  & 0.6472  & 0.6561  & 0.6607  & 0.5846  & 0.5986  & 0.5852  & 0.5703  \\
          & AutoInt & 0.6446  & 0.6396  & 0.6492  & 0.6509  & 0.5855  & 0.5995  & 0.5865  & 0.5701  \\
          & DCNv2 & 0.6532  & 0.6481  & 0.6576  & 0.6605  & 0.5835  & 0.5973  & 0.5839  & 0.5692  \\
    \midrule
    \multicolumn{1}{c}{\multirow{9}[2]{*}{\begin{sideways}Multi\end{sideways}}} & MMoE  & 0.6536  & 0.6488  & 0.6574  & 0.6615  & 0.5848  & 0.5988  & 0.5853  & 0.5714  \\
          & PLE   & 0.6505  & 0.6458  & 0.6547  & 0.6536  & 0.5851  & 0.5992  & 0.5858  & 0.5710  \\
          & STAR  & 0.6563  & 0.6498  & 0.6614  & 0.6603  & 0.5878  & 0.6007  & 0.5878  & 0.5622  \\
          & AdaSparse & 0.6572  & 0.6516  & 0.6605  & \underline{0.6696}  & 0.5855  & 0.5989  & 0.5855  & 0.5634  \\
          & HiNet & 0.6512  & 0.6467  & 0.6547  & 0.6583  & 0.5845  & 0.5989  & 0.5845  & 0.5592  \\
          & PEPNet (single) & 0.6566  & 0.6503  & 0.6613  & 0.6648  & 0.5905  & 0.6039  & 0.5922  & 0.5771  \\
          & PEPNet & 0.6552  & 0.6485  & 0.6601  & 0.6625  & 0.5849  & 0.5988  & 0.5856  & 0.5704  \\
          & ADL (group) & 0.6543  & 0.6488  & 0.6586  & 0.6623  & 0.5848  & 0.5987  & 0.5848  & 0.5666  \\
          & ADL   & 0.6469  & 0.6431  & 0.6506  & 0.6553  & 0.5829  & 0.5971  & 0.5829  & 0.5644  \\
    \midrule
    \multicolumn{1}{c}{\multirow{3}[2]{*}{\begin{sideways}Group\end{sideways}}} & Random & 0.6296  & 0.6264  & 0.6335  & 0.6295  & 0.5780  & 0.5891  & 0.5793  & 0.5600  \\
          & Manual & 0.6411  & 0.6375  & 0.6449  & 0.6414  &   -    &    -   &   -    & - \\
          & TAG   & 0.6523  & 0.6486  & 0.6560  & 0.6536  & 0.5889  & 0.6023  & 0.5899  & 0.5682  \\
    \midrule
          & CDC (end-to-end) & \underline{0.6592*}  & \textbf{0.6545* } & \underline{0.6631*}  & 0.6680  & \underline{0.5923*}  & \textbf{0.6065* } & \underline{0.5944*}  & \underline{0.5813}  \\
          & CDC   & \textbf{0.6613* } & \underline{0.6528*}  & \textbf{0.6664* } & \textbf{0.6728* } & \textbf{0.5972* } & \underline{0.6038}  & \textbf{0.5961* } & \textbf{0.5855* } \\
    \bottomrule
    \end{tabular}}%
  \label{tab:public_res}%
  \vspace{-11pt}
\end{table*}

\begin{table}[t]
  \centering
  \renewcommand{\arraystretch}{0.8}
  \caption{Performance of domain grouping methods with independently training of MDR models within each group.}
  \vspace{-8pt}
  \setlength{\tabcolsep}{2.2mm}
  {
    \footnotesize
    \begin{tabular}{cccccc}
    \toprule
    Dataset & Metric & Random & Manual & TAG   & CDC (split) \\
    \midrule
    \multirow{3}[2]{*}{Amazon} & DomainAUC & 0.6208  & 0.6359  & \underline{0.6576}  & \textbf{0.6660}*  \\
          & Major5AUC  & 0.6253  & 0.6406  & \underline{0.6615}  & \textbf{0.6707}*  \\
          & Minor10AUC  & 0.6280  & 0.6471  & \underline{0.6649}  & \textbf{0.6781}*  \\
    \midrule
    \multirow{3}[2]{*}{AliCCP} & DomainAUC & 0.5703  & -     & \underline{0.5805}  & \textbf{0.5906}*  \\
          & Major5AUC  & 0.5718  & -     & \underline{0.5813}  & \textbf{0.5922}*  \\
          & Minor10AUC  & 0.5505  & -     & \underline{0.5585}  & \textbf{0.5763}*  \\
    \bottomrule
    \end{tabular}}%
  \label{tab:separate}%
  \vspace{-12pt}
\end{table}%

\subsection{Metrics}

In the MDR setting, where item interactions are independently assessed within each domain, \textbf{DomainAUC} serves as the primary evaluation metric. This metric computes the AUC\cite{fawcett2006introduction} for each domain individually, averaging and weighting these values by their respective sample sizes. 
DomainAUC effectively reflects the model's ability to adapt across diverse domains, aligning well with potential online deployment gains. 
Given the limited space available for displaying all domain results, we introduce two supplementary metrics: \textbf{Major5AUC} and \textbf{Minor10AUC}. Major5AUC calculates the weighted average AUC of the five largest domains by sample size, while Minor10AUC focuses on the ten smallest domains. \textbf{TotalAUC} is provided as a global reference metric. In large-scale recommender systems, an improvement in AUC at the \textbf{0.001 level (1\textperthousand)} in CTR prediction tasks is considered significant and can lead to substantial commercial benefits\cite{song2019autoint, zhou2023hinet}.


\subsection{Implementation Details}
\label{sec:implement}
For optimization on public dataset, Adam\cite{kingma2014adam} is used with learning rates $[5e^{-4}, 1e^{-3}, 2e^{-3}, 3e^{-3}]$ and batch sizes $[1024, 2048, 4096]$ optimized through grid search. The number of clustering of "ADL (group)" and domain grouping methods is searched in $[3, 4, 5]$. The base model for domain grouping and CDC is selected from [MMoE, STAR, HiNet, PEPNet]. During the warm-up phase of CDC, the number of training batches is chosen from the range [1000, 2000]. During the lookahead updates, each update step involves updating only one batch. The initial domain count \(N\) for \(\mathcal{S}_{k,0}^t\) is set to 2. 

Three deployment versions of CDC have been developed. The first version, \textbf{CDC (end-to-end)}, implements a single-stage, end-to-end learning framework where the converged multi-domain model, as specified in Algorithm \ref{al:cdc}, is directly applied for inference. The second version, \textbf{CDC}, treats each previously generated target cluster $\mathcal{T}_k$ as a single domain and retrains a unified multi-domain model across these clusters. This approach offers the advantage of having similar time and storage costs as conventional MDR models, and often delivers better performance than the first version. This version has been selected for deployment in online A/B testing. The third version, \textbf{CDC (split)}, aligned with the setup in Table \ref{tab:separate} and Section \ref{sec:ablation}, trains independent models within each cluster $\mathcal{T}_k$ and evaluates them on their respective $\mathcal{S}_k$, potentially achieving the highest performance by minimizing domain conflicts. 
Notably, CDC (split) cannot measure TotalAUC, as the use of multiple models prevents global sorting of predictions across all domains.

\subsection{Performance on Public Datasets}


Table \ref{tab:public_res} and \ref{tab:separate} present the experimental results on two public datasets. From the results, we have the following observations:

(1) \textbf{CDC} and \textbf{CDC (end-to-end)} significantly outperform all the competitive baselines of all three types on both public datasets on all metrics. Specifically, compared to the best-performing baselines in Table \ref{tab:public_res}, CDC gains an average improvement of 4.14\textperthousand{} and 6.76\textperthousand{} on DomainAUC on the Amazon and AliCCP datasets, respectively.

\begin{table}[!t]
  \centering
  \renewcommand{\arraystretch}{0.8}
  \caption{Offline evaluation on the MDR-229M dataset.}
  \vspace{-9pt}
  \setlength{\tabcolsep}{4pt}
  {
    \footnotesize
    \begin{tabular}{lcccc}
    \toprule
    \multicolumn{1}{c}{Method} & DomainAUC & TotalAUC & Major5AUC & Minor20AUC \\
    \midrule
    DCN   & 0.00‰ & 0.00‰ & 0.00‰ & 0.00‰ \\
           STAR  & -3.33‰ & -0.63‰ & -2.79‰ & -5.12‰ \\
           PEPNet & +0.67‰ & \underline{+0.67‰} & +0.89‰ & +2.25‰ \\
    \midrule
    CDC-STAR & +0.38‰ & \textbf{+1.49‰} & +1.49‰ & -4.79‰ \\
           CDC-PEPNet & \underline{+1.89‰} & -4.89‰ & \underline{+2.37‰} & \underline{+2.46‰} \\
    \midrule
    CDC-$\mathcal{T}_k$ & +1.63‰ &    -   & +1.90‰ & -2.12‰ \\
           CDC-$\mathcal{D}$ & +2.19‰ &    -   & +2.15‰ & -0.77‰ \\
           CDC-$\mathcal{S}_k$   & \textbf{+2.85‰} &   -    & \textbf{+2.65‰} & \textbf{2.65‰} \\
    \bottomrule
    \end{tabular}%
  \label{tab:offline}}%
  \vspace{-15pt}
\end{table}%




(2) Compared to other multi-domain models, CDC (end-to-end) exhibits advantages by clustering domains with synergistic effects into the same group. This strategy mitigates negative transfer caused by indiscriminately blending all domains, and alleviates the data sparsity issue of minor domains. As for CDC, its improvements over other domain-grouping methods stem from the innovative design of adaptive domain affinity. Note that CDC (end-to-end) underperforms AdaSparse in the Minor10AUC metric on the Amazon dataset, possibly due to the shifting domain grouping during training, 
more significantly affecting minor domains.


(3) As shown in Table \ref{tab:separate}, \textbf{CDC (split)} achieve DomainAUC improvements of 8.46\textperthousand{} on Amazon and 10.15\textperthousand{} on AliCCP, outperforming all baselines. This significant enhancement highlights the importance of learning both training source domain sets $\{\mathcal{S}_k\}$ and target domain groups $\{\mathcal{T}_k\}$. 
Furthermore, the improvement over Table \ref{tab:public_res} on the Amazon dataset demonstrates that training on carefully selected subsets of domains, rather than all available domains, is more effective when significant inter-domain variability exists.

\subsection{Offline Evaluation}

For offline evaluation, we conduct representative model experiments on the industrial datasets to test the efficacy of CDC, given data scale and industrial platform constraints. The results on the MDR-229M dataset are shown in Table \ref{tab:offline}, using the Minor20AUC metric due to the large number of domains (over 80). The \textbf{DCN} single-domain model served as a benchmark to evaluate improvements of other models. \textbf{STAR} and \textbf{PEPNet} methods, which manually group domains based on business knowledge prior to training, contrast with \textbf{CDC-STAR} and \textbf{CDC-PEPNet} methods that employ CDC-derived grouping scheme to train a unified MDR model, respectively. These approaches demonstrate improvements over their respective base models, validating the efficacy of CDC's clustering outcomes. Methods including \textbf{CDC-$\mathcal{T}_k$}, \textbf{CDC-$\mathcal{D}$}, and \textbf{CDC-$\mathcal{S}_k$} each train independent PEPNet models within their clusters, using domain data sourced respectively from $\mathcal{T}_k$, all domains $\mathcal{D}$, and $\mathcal{S}_k$, with all evaluations and early stopping guided by $\mathcal{T}_k$. The notable performance of CDC-$\mathcal{S}_k$ highlights the importance of learning appropriate source training domain sets. Additionally, the prolonged experimental duration on \textbf{MDR-865M} limit the testing to \textbf{PEPNet} and \textbf{CDC-PEPNet}. The latter outperform the former with enhancements of $7.24$\textperthousand{} in DomainAUC and $7.09$\textperthousand{} in TotalAUC.


\begin{figure}[!t]
	\centering
    \vspace{-3pt}
    \includegraphics[height=0.10\textheight, trim=1 5 11 0,clip]{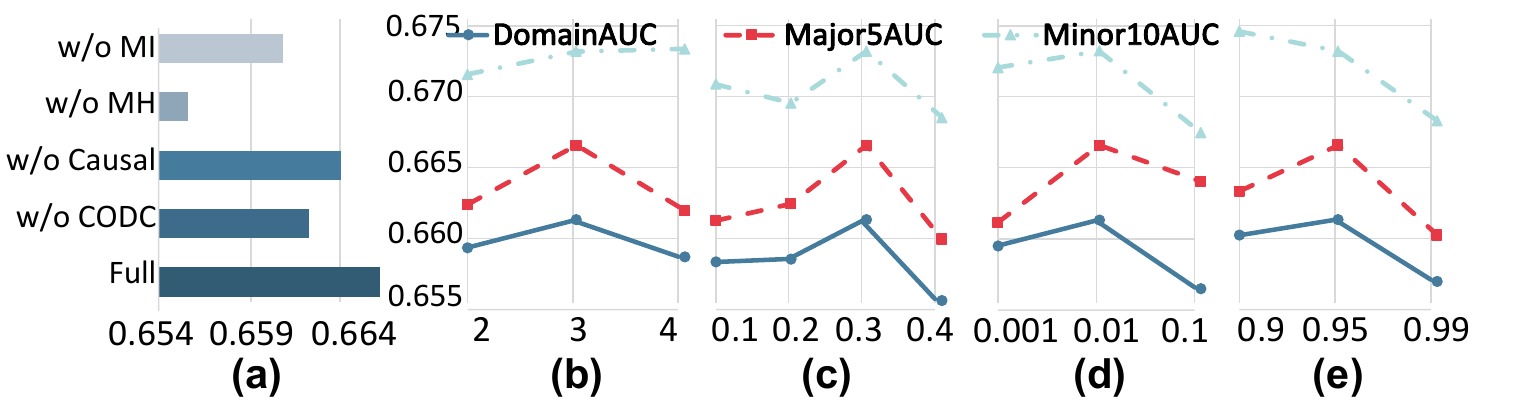}
	\centering
    \vspace{-10pt}
	\caption{Ablation study (subfigure (a)) and hyper-parameter analysis (subfigures (b)-(e) for \(K\), \(\alpha\), \(\rho_0\), and \(\beta\), respectively).}
    \vspace{-17pt}
	\label{fig:ablation_hyper}
\end{figure}

\subsection{Ablation Study}
\label{sec:ablation}

Figure \ref{fig:ablation_hyper}'s leftmost subplot presents the DomainAUC results for various ablation versions of our CDC on the Amazon dataset. Each domain cluster derived from CDC independently trains a multi-domain model, i.e., "CDC (split)", aligned with the setup in Table \ref{tab:separate}. We then evaluate the impact of removing key components. (1) CDC without the Isolated Domain Affinity Matrix $MI$ and the Hybrid Domain Affinity Matrix $MH$ are labeled as \textbf{w/o $MI$} and \textbf{w/o $MH$}, respectively. The results confirm the significant contribution of these matrices to capturing domain affinities, with $MH$ proving more crucial for enhancing high-order domain interactions. (2) Replacing the causal dependence contribution kernel $\kappa$ with a cosine function in Equation \ref{eq:dis} leads to the \textbf{w/o Causal} configuration. The decline in performance in this setup indicates that causal-based distances capture domain similarities more effectively than traditional cosine distances. (3) Without CODC, where we directly set $\mathcal{S}^t_{k,i}$ to $\mathcal{T}^t_{k,i}$ in each iteration, the model is referred to as \textbf{w/o CODC}. The diminished effectiveness of w/o CODC underscores the necessity of iteratively learning the training and target domain sets.

\subsection{Hyper-Parameter Study}
\label{sec:hyper}

In this section, we explore several key hyperparameters of the CDC mechanism on the Amazon dataset, including the number of clusters $K$, the smoothing factor $\alpha$ for history matrices in Equations (\ref{eq:mi}) and (\ref{eq:mh}), and the dynamic weighting coefficient $\rho^t = \rho^0 \cdot \beta^t$ for $P(d_u\mid k)$ in Equation (\ref{eq:JP}). Results are displayed in the rightmost four subfigures of Figure \ref{fig:ablation_hyper}, evaluating DomainAUC, Major5AUC, and Minor10AUC metrics. We vary the cluster number $K$ within $[2, 3, 4]$ and find that the optimal number of clusters for the 25 domains in the dataset is 3. $\alpha$ is tested over $[0.1, 0.2, 0.3, 0.4]$, with the best performance observed at $\alpha = 0.3$. This indicates that appropriately leveraging the historical affinity matrix can enhance the robustness and stability of the current matrix. Furthermore, we examine the parameters $\rho^0$ and $\beta$, with their optimal values determined to be $0.01$ and $0.95$, respectively. 
Combining results of these two parameters reveals that a minor weight $\rho^t$ for $P$ may slightly decrease performance; however, a larger $\rho^t$ typically results in significant performance degradation. 
This outcome likely stems from $P$, calculated based on the initial domain sets $\mathcal{S}_k^0$. This setup is primarily intended to offer global sight and mitigate the impact of inaccurate calculations of transfer gain $J$  during early optimization stages. As the optimization progresses and the reliability of $J$ improves,  an excessively high weight $\rho^t$ for $P$ can unnecessarily constrain the flexibility of clustering.

\subsection{Case Study}

\begin{figure}[!t]
	\centering
    \vspace{-8pt}
    \includegraphics[width=0.40\textwidth, trim=0 15 0 0,clip]{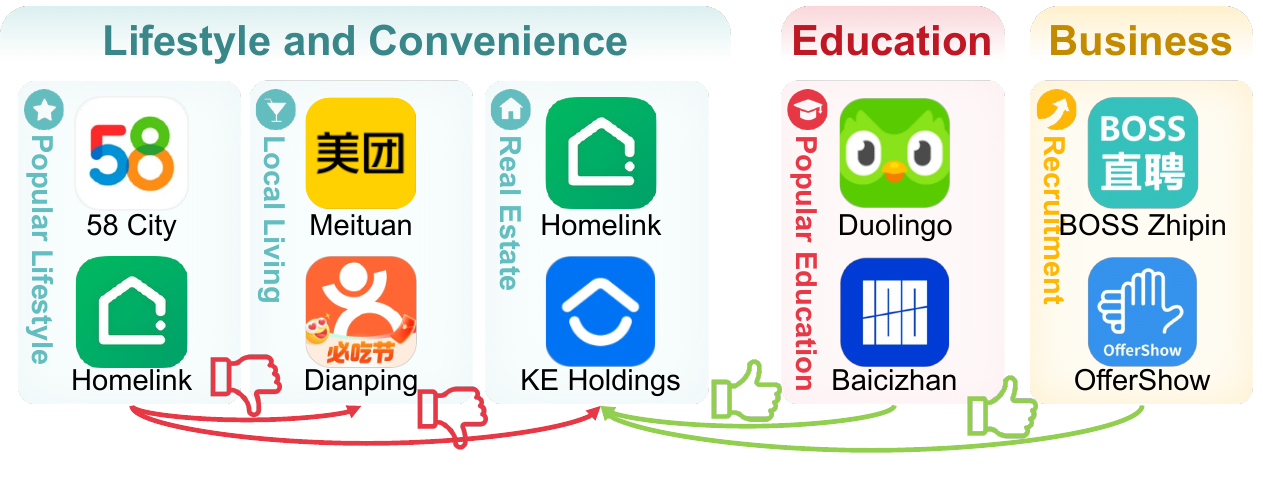}
	\centering
    \vspace{-10pt}
	\caption{Case study: Compared to manual grouping based on primary category, CDC reveals genuine cross-domain transfer effects to group domains that mutually benefit each other.}
    \vspace{-18pt}
	\label{fig:case}
\end{figure}

To demonstrate CDC’s industrial applicability, we present a case from our app marketplace. As shown in Figure \ref{fig:case}, users browse primary categories such as "Lifestyle and Convenience", "Education", and "Business", which further subdivide into secondary categories. Each secondary category's app exposure slate is treated as a separate domain. Traditionally, domains under the same primary category are manually grouped based on expert knowledge. However, CDC provides different insights. The Hybrid Domain Affinity Matrix \(MH\) shows that most domains negatively influence the "Real Estate" domain, except "Popular Education" and "Recruitment". As a result, CDC groups these three domains together, despite their differing primary categories. 

This grouping may seem counterintuitive, yet it reflects user behaviors. For example, housing searches often follow job or education changes. CDC captures these inter-domain synergies, providing insights into user decision-making that other grouping methods overlook, highlighting its practical value in industrial settings.

\subsection{Complexity Analysis}

We evaluate the scalability of the proposed CDC method by comparing its parameter storage and training time with other models on the public Amazon dataset and the industrial MDR-229M dataset (Table \ref{tab:complexity}). For Amazon, the "CDC (end-to-end)" configuration incurs an 8.9\% increase in training time, while the MDR-229M shows a 1.5\% increase. This increase remains stable across different sizes of interaction data. The complexity of the "CDC" configuration aligns with other baseline models, and this is the version deployed online.

\begin{table}[!t]
  \centering  
  \renewcommand{\arraystretch}{0.8}
  \vspace{-9pt}
  \caption{Average time and space complexity of CDC and its competitors on the Amazon and MDR-229M datasets.}
  \vspace{-8pt}
  \setlength{\tabcolsep}{1pt}
  {  
    \footnotesize
    \begin{tabular}{lcccc}
    \toprule
    \multicolumn{1}{c}{\multirow{2}[0]{*}{Methods}} & \multicolumn{2}{c}{Amazon} & \multicolumn{2}{c}{MDR-229M} \\
          & Storage (MB) & Training (min) & Storage (rel) & Training (rel) \\
    \midrule
    Single & 1969.14  & 162.36  & 1.00x  & 1.00x  \\
    Multi & 1980.71  & 214.95  & OOM   & OOM \\
    Grouping & 1971.90  & 387.91+128.17 & 1.00x  & 1.13x  \\
    CDC (end-to-end) & 1969.06  & 234.15  & 1.00x & 1.25x \\
    CDC   & 1971.90  & 139.32  & 1.00x  & 1.09x  \\
    \bottomrule
    \end{tabular}}
  \label{tab:complexity}
  \vspace{-18pt}
\end{table}%

\section{Online Evaluation}


We conducted online A/B testing experiments over a 14-day period within our app marketplace's recommender system, \textbf{covering 64 domains and 20\% of user traffic}. This translates to \textbf{millions of users and billions of exposures daily}. Utilizing the domain clustering results $\{\mathcal{T}_k\}_{k=1}^K$ derived from our CDC, we trained a unified multi-domain model with the same architecture as the fully-deployed baseline. This implementation matched the training costs of the current baseline and resulted in a significant increase in overall revenue metric eCPM, or effective cost per mille, by $4.9\%$. Notably, it improved eCPM in $56$ of these domains, demonstrating its effectiveness in managing a large number of domains in industry.

\section{Conclusion}



We propose a novel CDC framework to address domain clustering challenges in multi-domain recommendation systems with a large number of domains. We use two affinity matrices to dynamically capture inter-domain transfer effects in different transfer patterns. We then apply causal discovery to adaptively integrate these matrices. Finally, the CODC method iteratively optimizes target domain clusters and their corresponding training source domain sets. Extensive evaluations on public datasets and real-world recommendation systems demonstrate that CDC significantly outperforms existing multi-domain strategies and domain grouping methods.

\begin{acks}
Supported by the National Key Research and Development Program of China under Grant No. 2024YFF0729003, the National Natural Science Foundation of China under Grant Nos. 62176014 and 62276015, and the Fundamental Research Funds for the Central Universities.
\end{acks}

\bibliographystyle{ACM-Reference-Format}
\bibliography{ref}


\begin{thebibliography}{41}


\ifx \showCODEN    \undefined \def \showCODEN     #1{\unskip}     \fi
\ifx \showISBNx    \undefined \def \showISBNx     #1{\unskip}     \fi
\ifx \showISBNxiii \undefined \def \showISBNxiii  #1{\unskip}     \fi
\ifx \showISSN     \undefined \def \showISSN      #1{\unskip}     \fi
\ifx \showLCCN     \undefined \def \showLCCN      #1{\unskip}     \fi
\ifx \shownote     \undefined \def \shownote      #1{#1}          \fi
\ifx \showarticletitle \undefined \def \showarticletitle #1{#1}   \fi
\ifx \showURL      \undefined \def \showURL       {\relax}        \fi
\providecommand\bibfield[2]{#2}
\providecommand\bibinfo[2]{#2}
\providecommand\natexlab[1]{#1}
\providecommand\showeprint[2][]{arXiv:#2}

\bibitem[Bai and Zhao(2022)]%
        {bai2022saliency}
\bibfield{author}{\bibinfo{person}{Guangji Bai} {and} \bibinfo{person}{Liang Zhao}.} \bibinfo{year}{2022}\natexlab{}.
\newblock \showarticletitle{Saliency-Regularized Deep Multi-Task Learning}. In \bibinfo{booktitle}{\emph{Proceedings of the 28th ACM SIGKDD Conference on Knowledge Discovery and Data Mining}} (Washington DC, USA) \emph{(\bibinfo{series}{KDD '22})}. \bibinfo{publisher}{Association for Computing Machinery}, \bibinfo{address}{New York, NY, USA}, \bibinfo{pages}{15–25}.
\newblock
\showISBNx{9781450393850}
\href{https://doi.org/10.1145/3534678.3539442}{doi:\nolinkurl{10.1145/3534678.3539442}}


\bibitem[Chang et~al\mbox{.}(2023)]%
        {chang2023pepnet}
\bibfield{author}{\bibinfo{person}{Jianxin Chang}, \bibinfo{person}{Chenbin Zhang}, \bibinfo{person}{Yiqun Hui}, \bibinfo{person}{Dewei Leng}, \bibinfo{person}{Yanan Niu}, \bibinfo{person}{Yang Song}, {and} \bibinfo{person}{Kun Gai}.} \bibinfo{year}{2023}\natexlab{}.
\newblock \showarticletitle{Pepnet: Parameter and embedding personalized network for infusing with personalized prior information}. In \bibinfo{booktitle}{\emph{Proceedings of the 29th ACM SIGKDD Conference on Knowledge Discovery and Data Mining}}. \bibinfo{pages}{3795--3804}.
\newblock


\bibitem[Dhillon et~al\mbox{.}(2004)]%
        {dhillon2004kernel}
\bibfield{author}{\bibinfo{person}{Inderjit~S Dhillon}, \bibinfo{person}{Yuqiang Guan}, {and} \bibinfo{person}{Brian Kulis}.} \bibinfo{year}{2004}\natexlab{}.
\newblock \showarticletitle{Kernel k-means: spectral clustering and normalized cuts}. In \bibinfo{booktitle}{\emph{Proceedings of the tenth ACM SIGKDD international conference on Knowledge discovery and data mining}}. \bibinfo{pages}{551--556}.
\newblock


\bibitem[Dhillon et~al\mbox{.}(2003)]%
        {dhillon2003diametrical}
\bibfield{author}{\bibinfo{person}{Inderjit~S Dhillon}, \bibinfo{person}{Edward~M Marcotte}, {and} \bibinfo{person}{Usman Roshan}.} \bibinfo{year}{2003}\natexlab{}.
\newblock \showarticletitle{Diametrical clustering for identifying anti-correlated gene clusters}.
\newblock \bibinfo{journal}{\emph{Bioinformatics}} \bibinfo{volume}{19}, \bibinfo{number}{13} (\bibinfo{year}{2003}), \bibinfo{pages}{1612--1619}.
\newblock


\bibitem[Fawcett(2006)]%
        {fawcett2006introduction}
\bibfield{author}{\bibinfo{person}{Tom Fawcett}.} \bibinfo{year}{2006}\natexlab{}.
\newblock \showarticletitle{An introduction to ROC analysis}.
\newblock \bibinfo{journal}{\emph{Pattern recognition letters}} \bibinfo{volume}{27}, \bibinfo{number}{8} (\bibinfo{year}{2006}), \bibinfo{pages}{861--874}.
\newblock


\bibitem[Fifty et~al\mbox{.}(2021)]%
        {fifty2021efficiently}
\bibfield{author}{\bibinfo{person}{Chris Fifty}, \bibinfo{person}{Ehsan Amid}, \bibinfo{person}{Zhe Zhao}, \bibinfo{person}{Tianhe Yu}, \bibinfo{person}{Rohan Anil}, {and} \bibinfo{person}{Chelsea Finn}.} \bibinfo{year}{2021}\natexlab{}.
\newblock \showarticletitle{Efficiently Identifying Task Groupings for Multi-Task Learning}. In \bibinfo{booktitle}{\emph{Advances in Neural Information Processing Systems}}, \bibfield{editor}{\bibinfo{person}{M.~Ranzato}, \bibinfo{person}{A.~Beygelzimer}, \bibinfo{person}{Y.~Dauphin}, \bibinfo{person}{P.S. Liang}, {and} \bibinfo{person}{J.~Wortman Vaughan}} (Eds.), Vol.~\bibinfo{volume}{34}. \bibinfo{publisher}{Curran Associates, Inc.}, \bibinfo{pages}{27503--27516}.
\newblock
\urldef\tempurl%
\url{https://proceedings.neurips.cc/paper_files/paper/2021/file/e77910ebb93b511588557806310f78f1-Paper.pdf}
\showURL{%
\tempurl}


\bibitem[Guo et~al\mbox{.}(2017)]%
        {guo2017deepfm}
\bibfield{author}{\bibinfo{person}{Huifeng Guo}, \bibinfo{person}{Ruiming Tang}, \bibinfo{person}{Yunming Ye}, \bibinfo{person}{Zhenguo Li}, {and} \bibinfo{person}{Xiuqiang He}.} \bibinfo{year}{2017}\natexlab{}.
\newblock \showarticletitle{DeepFM: a factorization-machine based neural network for CTR prediction}. In \bibinfo{booktitle}{\emph{Proceedings of the 26th International Joint Conference on Artificial Intelligence}}. \bibinfo{pages}{1725--1731}.
\newblock


\bibitem[Ha et~al\mbox{.}(2016)]%
        {ha2016hypernetworks}
\bibfield{author}{\bibinfo{person}{David Ha}, \bibinfo{person}{Andrew Dai}, {and} \bibinfo{person}{Quoc~V Le}.} \bibinfo{year}{2016}\natexlab{}.
\newblock \showarticletitle{Hypernetworks}.
\newblock \bibinfo{journal}{\emph{arXiv preprint arXiv:1609.09106}} (\bibinfo{year}{2016}).
\newblock


\bibitem[Iyer et~al\mbox{.}(1999)]%
        {iyer1999transcriptional}
\bibfield{author}{\bibinfo{person}{Vishwanath~R Iyer}, \bibinfo{person}{Michael~B Eisen}, \bibinfo{person}{Douglas~T Ross}, \bibinfo{person}{Greg Schuler}, \bibinfo{person}{Troy Moore}, \bibinfo{person}{Jeffrey~CF Lee}, \bibinfo{person}{Jeffrey~M Trent}, \bibinfo{person}{Louis~M Staudt}, \bibinfo{person}{James Hudson~Jr}, \bibinfo{person}{Mark~S Boguski}, {et~al\mbox{.}}} \bibinfo{year}{1999}\natexlab{}.
\newblock \showarticletitle{The transcriptional program in the response of human fibroblasts to serum}.
\newblock \bibinfo{journal}{\emph{science}} \bibinfo{volume}{283}, \bibinfo{number}{5398} (\bibinfo{year}{1999}), \bibinfo{pages}{83--87}.
\newblock


\bibitem[Jia et~al\mbox{.}(2024)]%
        {jia2024d3}
\bibfield{author}{\bibinfo{person}{Pengyue Jia}, \bibinfo{person}{Yichao Wang}, \bibinfo{person}{Shanru Lin}, \bibinfo{person}{Xiaopeng Li}, \bibinfo{person}{Xiangyu Zhao}, \bibinfo{person}{Huifeng Guo}, {and} \bibinfo{person}{Ruiming Tang}.} \bibinfo{year}{2024}\natexlab{}.
\newblock \showarticletitle{D3: A Methodological Exploration of Domain Division, Modeling, and Balance in Multi-Domain Recommendations}. In \bibinfo{booktitle}{\emph{Proceedings of the AAAI Conference on Artificial Intelligence}}, Vol.~\bibinfo{volume}{38}. \bibinfo{pages}{8553--8561}.
\newblock


\bibitem[Kingma and Ba(2015)]%
        {kingma2014adam}
\bibfield{author}{\bibinfo{person}{Diederik~P. Kingma} {and} \bibinfo{person}{Jimmy Ba}.} \bibinfo{year}{2015}\natexlab{}.
\newblock \showarticletitle{Adam: {A} Method for Stochastic Optimization}. In \bibinfo{booktitle}{\emph{3rd International Conference on Learning Representations, {ICLR} 2015, San Diego, CA, USA, May 7-9, 2015, Conference Track Proceedings}}, \bibfield{editor}{\bibinfo{person}{Yoshua Bengio} {and} \bibinfo{person}{Yann LeCun}} (Eds.).
\newblock
\urldef\tempurl%
\url{http://arxiv.org/abs/1412.6980}
\showURL{%
\tempurl}


\bibitem[Li et~al\mbox{.}(2023)]%
        {li2023adl}
\bibfield{author}{\bibinfo{person}{Jinyun Li}, \bibinfo{person}{Huiwen Zheng}, \bibinfo{person}{Yuanlin Liu}, \bibinfo{person}{Minfang Lu}, \bibinfo{person}{Lixia Wu}, {and} \bibinfo{person}{Haoyuan Hu}.} \bibinfo{year}{2023}\natexlab{}.
\newblock \showarticletitle{ADL: Adaptive Distribution Learning Framework for Multi-Scenario CTR Prediction}. In \bibinfo{booktitle}{\emph{Proceedings of the 46th International ACM SIGIR Conference on Research and Development in Information Retrieval}}. \bibinfo{pages}{1786--1790}.
\newblock


\bibitem[Liu et~al\mbox{.}(2023a)]%
        {liu2023multitask}
\bibfield{author}{\bibinfo{person}{Qingyun Liu}, \bibinfo{person}{Zhe Zhao}, \bibinfo{person}{Liang Liu}, \bibinfo{person}{Zhen Zhang}, \bibinfo{person}{Junjie Shan}, \bibinfo{person}{Yuening Li}, \bibinfo{person}{Shuchao Bi}, \bibinfo{person}{Lichan Hong}, {and} \bibinfo{person}{Ed~H Chi}.} \bibinfo{year}{2023}\natexlab{a}.
\newblock \showarticletitle{Multitask Ranking System for Immersive Feed and No More Clicks: A Case Study of Short-Form Video Recommendation}. In \bibinfo{booktitle}{\emph{Proceedings of the 32nd ACM International Conference on Information and Knowledge Management}}. \bibinfo{pages}{4709--4716}.
\newblock


\bibitem[Liu et~al\mbox{.}(2023b)]%
        {liu2023dtrn}
\bibfield{author}{\bibinfo{person}{Qi Liu}, \bibinfo{person}{Zhilong Zhou}, \bibinfo{person}{Gangwei Jiang}, \bibinfo{person}{Tiezheng Ge}, {and} \bibinfo{person}{Defu Lian}.} \bibinfo{year}{2023}\natexlab{b}.
\newblock \showarticletitle{Deep task-specific bottom representation network for multi-task recommendation}. In \bibinfo{booktitle}{\emph{Proceedings of the 32nd ACM International Conference on Information and Knowledge Management}}. \bibinfo{pages}{1637--1646}.
\newblock


\bibitem[Liu et~al\mbox{.}(2022)]%
        {liu2022autolambda}
\bibfield{author}{\bibinfo{person}{Shikun Liu}, \bibinfo{person}{Stephen James}, \bibinfo{person}{Andrew~J Davison}, {and} \bibinfo{person}{Edward Johns}.} \bibinfo{year}{2022}\natexlab{}.
\newblock \showarticletitle{Auto-Lambda: Disentangling Dynamic Task Relationships}.
\newblock \bibinfo{journal}{\emph{Transactions on Machine Learning Research}} (\bibinfo{year}{2022}).
\newblock


\bibitem[Liu et~al\mbox{.}(2019)]%
        {liu2019loss}
\bibfield{author}{\bibinfo{person}{Shengchao Liu}, \bibinfo{person}{Yingyu Liang}, {and} \bibinfo{person}{Anthony Gitter}.} \bibinfo{year}{2019}\natexlab{}.
\newblock \showarticletitle{Loss-balanced task weighting to reduce negative transfer in multi-task learning}. In \bibinfo{booktitle}{\emph{Proceedings of the AAAI conference on artificial intelligence}}, Vol.~\bibinfo{volume}{33}. \bibinfo{pages}{9977--9978}.
\newblock


\bibitem[Liu(2015)]%
        {liu2015reverse}
\bibfield{author}{\bibinfo{person}{Zhi-Ping Liu}.} \bibinfo{year}{2015}\natexlab{}.
\newblock \showarticletitle{Reverse engineering of genome-wide gene regulatory networks from gene expression data}.
\newblock \bibinfo{journal}{\emph{Current genomics}} \bibinfo{volume}{16}, \bibinfo{number}{1} (\bibinfo{year}{2015}), \bibinfo{pages}{3--22}.
\newblock


\bibitem[Luo et~al\mbox{.}(2024)]%
        {luo2024ci4rs}
\bibfield{author}{\bibinfo{person}{Huishi Luo}, \bibinfo{person}{Fuzhen Zhuang}, \bibinfo{person}{Ruobing Xie}, \bibinfo{person}{Hengshu Zhu}, \bibinfo{person}{Deqing Wang}, \bibinfo{person}{Zhulin An}, {and} \bibinfo{person}{Yongjun Xu}.} \bibinfo{year}{2024}\natexlab{}.
\newblock \showarticletitle{A survey on causal inference for recommendation}.
\newblock \bibinfo{journal}{\emph{The Innovation}} \bibinfo{volume}{5}, \bibinfo{number}{2} (\bibinfo{year}{2024}), \bibinfo{pages}{100590}.
\newblock
\showISSN{2666-6758}
\href{https://doi.org/10.1016/j.xinn.2024.100590}{doi:\nolinkurl{10.1016/j.xinn.2024.100590}}


\bibitem[Ma et~al\mbox{.}(2018b)]%
        {ma2018mmoe}
\bibfield{author}{\bibinfo{person}{Jiaqi Ma}, \bibinfo{person}{Zhe Zhao}, \bibinfo{person}{Xinyang Yi}, \bibinfo{person}{Jilin Chen}, \bibinfo{person}{Lichan Hong}, {and} \bibinfo{person}{Ed~H Chi}.} \bibinfo{year}{2018}\natexlab{b}.
\newblock \showarticletitle{Modeling task relationships in multi-task learning with multi-gate mixture-of-experts}. In \bibinfo{booktitle}{\emph{Proceedings of the 24th ACM SIGKDD international conference on knowledge discovery \& data mining}}. \bibinfo{pages}{1930--1939}.
\newblock


\bibitem[Ma et~al\mbox{.}(2018a)]%
        {ma2018entire}
\bibfield{author}{\bibinfo{person}{Xiao Ma}, \bibinfo{person}{Liqin Zhao}, \bibinfo{person}{Guan Huang}, \bibinfo{person}{Zhi Wang}, \bibinfo{person}{Zelin Hu}, \bibinfo{person}{Xiaoqiang Zhu}, {and} \bibinfo{person}{Kun Gai}.} \bibinfo{year}{2018}\natexlab{a}.
\newblock \showarticletitle{Entire space multi-task model: An effective approach for estimating post-click conversion rate}. In \bibinfo{booktitle}{\emph{The 41st International ACM SIGIR Conference on Research \& Development in Information Retrieval}}. \bibinfo{pages}{1137--1140}.
\newblock


\bibitem[Markham et~al\mbox{.}(2022)]%
        {markham2022distance}
\bibfield{author}{\bibinfo{person}{Alex Markham}, \bibinfo{person}{Richeek Das}, {and} \bibinfo{person}{Moritz Grosse-Wentrup}.} \bibinfo{year}{2022}\natexlab{}.
\newblock \showarticletitle{A distance covariance-based kernel for nonlinear causal clustering in heterogeneous populations}. In \bibinfo{booktitle}{\emph{Conference on Causal Learning and Reasoning}}. PMLR, \bibinfo{pages}{542--558}.
\newblock


\bibitem[Ni et~al\mbox{.}(2019)]%
        {ni2019justifying}
\bibfield{author}{\bibinfo{person}{Jianmo Ni}, \bibinfo{person}{Jiacheng Li}, {and} \bibinfo{person}{Julian McAuley}.} \bibinfo{year}{2019}\natexlab{}.
\newblock \showarticletitle{Justifying recommendations using distantly-labeled reviews and fine-grained aspects}. In \bibinfo{booktitle}{\emph{Proceedings of the 2019 conference on empirical methods in natural language processing and the 9th international joint conference on natural language processing (EMNLP-IJCNLP)}}. \bibinfo{pages}{188--197}.
\newblock


\bibitem[Pearl(2009)]%
        {pearl2009causality}
\bibfield{author}{\bibinfo{person}{Judea Pearl}.} \bibinfo{year}{2009}\natexlab{}.
\newblock \bibinfo{booktitle}{\emph{Causality} (\bibinfo{edition}{2} ed.)}.
\newblock \bibinfo{publisher}{Cambridge University Press}.
\newblock
\href{https://doi.org/10.1017/CBO9780511803161}{doi:\nolinkurl{10.1017/CBO9780511803161}}


\bibitem[Raychaudhuri et~al\mbox{.}(2022)]%
        {raychaudhuri2022controllable}
\bibfield{author}{\bibinfo{person}{Dripta~S Raychaudhuri}, \bibinfo{person}{Yumin Suh}, \bibinfo{person}{Samuel Schulter}, \bibinfo{person}{Xiang Yu}, \bibinfo{person}{Masoud Faraki}, \bibinfo{person}{Amit~K Roy-Chowdhury}, {and} \bibinfo{person}{Manmohan Chandraker}.} \bibinfo{year}{2022}\natexlab{}.
\newblock \showarticletitle{Controllable dynamic multi-task architectures}. In \bibinfo{booktitle}{\emph{Proceedings of the IEEE/CVF Conference on Computer Vision and Pattern Recognition}}. \bibinfo{pages}{10955--10964}.
\newblock


\bibitem[Shen et~al\mbox{.}(2021)]%
        {shen2021sarnet}
\bibfield{author}{\bibinfo{person}{Qijie Shen}, \bibinfo{person}{Wanjie Tao}, \bibinfo{person}{Jing Zhang}, \bibinfo{person}{Hong Wen}, \bibinfo{person}{Zulong Chen}, {and} \bibinfo{person}{Quan Lu}.} \bibinfo{year}{2021}\natexlab{}.
\newblock \showarticletitle{SAR-Net: A scenario-aware ranking network for personalized fair recommendation in hundreds of travel scenarios}. In \bibinfo{booktitle}{\emph{Proceedings of the 30th ACM International Conference on Information \& Knowledge Management}}. \bibinfo{pages}{4094--4103}.
\newblock


\bibitem[Sheng et~al\mbox{.}(2021)]%
        {sheng2021star}
\bibfield{author}{\bibinfo{person}{Xiang-Rong Sheng}, \bibinfo{person}{Liqin Zhao}, \bibinfo{person}{Guorui Zhou}, \bibinfo{person}{Xinyao Ding}, \bibinfo{person}{Binding Dai}, \bibinfo{person}{Qiang Luo}, \bibinfo{person}{Siran Yang}, \bibinfo{person}{Jingshan Lv}, \bibinfo{person}{Chi Zhang}, \bibinfo{person}{Hongbo Deng}, {et~al\mbox{.}}} \bibinfo{year}{2021}\natexlab{}.
\newblock \showarticletitle{One model to serve all: Star topology adaptive recommender for multi-domain ctr prediction}. In \bibinfo{booktitle}{\emph{Proceedings of the 30th ACM International Conference on Information \& Knowledge Management}}. \bibinfo{pages}{4104--4113}.
\newblock


\bibitem[Sherif et~al\mbox{.}(2024)]%
        {sherif2024stg}
\bibfield{author}{\bibinfo{person}{Ammar Sherif}, \bibinfo{person}{Abubakar Abid}, \bibinfo{person}{Mustafa Elattar}, {and} \bibinfo{person}{Mohamed ElHelw}.} \bibinfo{year}{2024}\natexlab{}.
\newblock \showarticletitle{STG-MTL: scalable task grouping for multi-task learning using data maps}.
\newblock \bibinfo{journal}{\emph{Machine Learning: Science and Technology}} \bibinfo{volume}{5}, \bibinfo{number}{2} (\bibinfo{year}{2024}), \bibinfo{pages}{025068}.
\newblock


\bibitem[Song et~al\mbox{.}(2019)]%
        {song2019autoint}
\bibfield{author}{\bibinfo{person}{Weiping Song}, \bibinfo{person}{Chence Shi}, \bibinfo{person}{Zhiping Xiao}, \bibinfo{person}{Zhijian Duan}, \bibinfo{person}{Yewen Xu}, \bibinfo{person}{Ming Zhang}, {and} \bibinfo{person}{Jian Tang}.} \bibinfo{year}{2019}\natexlab{}.
\newblock \showarticletitle{Autoint: Automatic feature interaction learning via self-attentive neural networks}. In \bibinfo{booktitle}{\emph{Proceedings of the 28th ACM international conference on information and knowledge management}}. \bibinfo{pages}{1161--1170}.
\newblock


\bibitem[Song et~al\mbox{.}(2022)]%
        {song2022efficient}
\bibfield{author}{\bibinfo{person}{Xiaozhuang Song}, \bibinfo{person}{Shun Zheng}, \bibinfo{person}{Wei Cao}, \bibinfo{person}{James Yu}, {and} \bibinfo{person}{Jiang Bian}.} \bibinfo{year}{2022}\natexlab{}.
\newblock \showarticletitle{Efficient and effective multi-task grouping via meta learning on task combinations}.
\newblock \bibinfo{journal}{\emph{Advances in Neural Information Processing Systems}}  \bibinfo{volume}{35} (\bibinfo{year}{2022}), \bibinfo{pages}{37647--37659}.
\newblock


\bibitem[Standley et~al\mbox{.}(2020)]%
        {standley2020tasks}
\bibfield{author}{\bibinfo{person}{Trevor Standley}, \bibinfo{person}{Amir Zamir}, \bibinfo{person}{Dawn Chen}, \bibinfo{person}{Leonidas Guibas}, \bibinfo{person}{Jitendra Malik}, {and} \bibinfo{person}{Silvio Savarese}.} \bibinfo{year}{2020}\natexlab{}.
\newblock \showarticletitle{Which tasks should be learned together in multi-task learning?}. In \bibinfo{booktitle}{\emph{International conference on machine learning}}. PMLR, \bibinfo{pages}{9120--9132}.
\newblock


\bibitem[Swayamdipta et~al\mbox{.}(2020)]%
        {swayamdipta2020dataset}
\bibfield{author}{\bibinfo{person}{Swabha Swayamdipta}, \bibinfo{person}{Roy Schwartz}, \bibinfo{person}{Nicholas Lourie}, \bibinfo{person}{Yizhong Wang}, \bibinfo{person}{Hannaneh Hajishirzi}, \bibinfo{person}{Noah~A Smith}, {and} \bibinfo{person}{Yejin Choi}.} \bibinfo{year}{2020}\natexlab{}.
\newblock \showarticletitle{Dataset cartography: Mapping and diagnosing datasets with training dynamics}.
\newblock \bibinfo{journal}{\emph{arXiv preprint arXiv:2009.10795}} (\bibinfo{year}{2020}).
\newblock


\bibitem[Tang et~al\mbox{.}(2020)]%
        {tang2020ple}
\bibfield{author}{\bibinfo{person}{Hongyan Tang}, \bibinfo{person}{Junning Liu}, \bibinfo{person}{Ming Zhao}, {and} \bibinfo{person}{Xudong Gong}.} \bibinfo{year}{2020}\natexlab{}.
\newblock \showarticletitle{Progressive layered extraction (ple): A novel multi-task learning (mtl) model for personalized recommendations}. In \bibinfo{booktitle}{\emph{Proceedings of the 14th ACM Conference on Recommender Systems}}. \bibinfo{pages}{269--278}.
\newblock


\bibitem[Wang et~al\mbox{.}(2024)]%
        {wang2024principled}
\bibfield{author}{\bibinfo{person}{Chenguang Wang}, \bibinfo{person}{Xuanhao Pan}, {and} \bibinfo{person}{Tianshu Yu}.} \bibinfo{year}{2024}\natexlab{}.
\newblock \showarticletitle{Towards Principled Task Grouping for Multi-Task Learning}.
\newblock \bibinfo{journal}{\emph{arXiv preprint arXiv:2402.15328}} (\bibinfo{year}{2024}).
\newblock


\bibitem[Wang et~al\mbox{.}(2017)]%
        {wang2017dcn}
\bibfield{author}{\bibinfo{person}{Ruoxi Wang}, \bibinfo{person}{Bin Fu}, \bibinfo{person}{Gang Fu}, {and} \bibinfo{person}{Mingliang Wang}.} \bibinfo{year}{2017}\natexlab{}.
\newblock \showarticletitle{Deep \& cross network for ad click predictions}.
\newblock In \bibinfo{booktitle}{\emph{Proceedings of the ADKDD'17}}. \bibinfo{pages}{1--7}.
\newblock


\bibitem[Wang et~al\mbox{.}(2021)]%
        {wang2021dcnv2}
\bibfield{author}{\bibinfo{person}{Ruoxi Wang}, \bibinfo{person}{Rakesh Shivanna}, \bibinfo{person}{Derek Cheng}, \bibinfo{person}{Sagar Jain}, \bibinfo{person}{Dong Lin}, \bibinfo{person}{Lichan Hong}, {and} \bibinfo{person}{Ed Chi}.} \bibinfo{year}{2021}\natexlab{}.
\newblock \showarticletitle{Dcn v2: Improved deep \& cross network and practical lessons for web-scale learning to rank systems}. In \bibinfo{booktitle}{\emph{Proceedings of the web conference 2021}}. \bibinfo{pages}{1785--1797}.
\newblock


\bibitem[Wang et~al\mbox{.}(2022)]%
        {wang2022causalint}
\bibfield{author}{\bibinfo{person}{Yichao Wang}, \bibinfo{person}{Huifeng Guo}, \bibinfo{person}{Bo Chen}, \bibinfo{person}{Weiwen Liu}, \bibinfo{person}{Zhirong Liu}, \bibinfo{person}{Qi Zhang}, \bibinfo{person}{Zhicheng He}, \bibinfo{person}{Hongkun Zheng}, \bibinfo{person}{Weiwei Yao}, \bibinfo{person}{Muyu Zhang}, {et~al\mbox{.}}} \bibinfo{year}{2022}\natexlab{}.
\newblock \showarticletitle{Causalint: Causal inspired intervention for multi-scenario recommendation}. In \bibinfo{booktitle}{\emph{Proceedings of the 28th ACM SIGKDD Conference on Knowledge Discovery and Data Mining}}. \bibinfo{pages}{4090--4099}.
\newblock


\bibitem[Yang et~al\mbox{.}(2022)]%
        {yang2022adasparse}
\bibfield{author}{\bibinfo{person}{Xuanhua Yang}, \bibinfo{person}{Xiaoyu Peng}, \bibinfo{person}{Penghui Wei}, \bibinfo{person}{Shaoguo Liu}, \bibinfo{person}{Liang Wang}, {and} \bibinfo{person}{Bo Zheng}.} \bibinfo{year}{2022}\natexlab{}.
\newblock \showarticletitle{Adasparse: Learning adaptively sparse structures for multi-domain click-through rate prediction}. In \bibinfo{booktitle}{\emph{Proceedings of the 31st ACM International Conference on Information \& Knowledge Management}}. \bibinfo{pages}{4635--4639}.
\newblock


\bibitem[Zhang et~al\mbox{.}(2022)]%
        {zhang2022sass}
\bibfield{author}{\bibinfo{person}{Yuanliang Zhang}, \bibinfo{person}{Xiaofeng Wang}, \bibinfo{person}{Jinxin Hu}, \bibinfo{person}{Ke Gao}, \bibinfo{person}{Chenyi Lei}, {and} \bibinfo{person}{Fei Fang}.} \bibinfo{year}{2022}\natexlab{}.
\newblock \showarticletitle{Scenario-adaptive and self-supervised model for multi-scenario personalized recommendation}. In \bibinfo{booktitle}{\emph{Proceedings of the 31st ACM International Conference on Information \& Knowledge Management}}. \bibinfo{pages}{3674--3683}.
\newblock


\bibitem[Zhou et~al\mbox{.}(2023)]%
        {zhou2023hinet}
\bibfield{author}{\bibinfo{person}{Jie Zhou}, \bibinfo{person}{Xianshuai Cao}, \bibinfo{person}{Wenhao Li}, \bibinfo{person}{Lin Bo}, \bibinfo{person}{Kun Zhang}, \bibinfo{person}{Chuan Luo}, {and} \bibinfo{person}{Qian Yu}.} \bibinfo{year}{2023}\natexlab{}.
\newblock \showarticletitle{Hinet: Novel multi-scenario \& multi-task learning with hierarchical information extraction}. In \bibinfo{booktitle}{\emph{2023 IEEE 39th International Conference on Data Engineering (ICDE)}}. IEEE, \bibinfo{pages}{2969--2975}.
\newblock


\bibitem[Zhuang et~al\mbox{.}(2020)]%
        {zhuang2020survey}
\bibfield{author}{\bibinfo{person}{Fuzhen Zhuang}, \bibinfo{person}{Zhiyuan Qi}, \bibinfo{person}{Keyu Duan}, \bibinfo{person}{Dongbo Xi}, \bibinfo{person}{Yongchun Zhu}, \bibinfo{person}{Hengshu Zhu}, \bibinfo{person}{Hui Xiong}, {and} \bibinfo{person}{Qing He}.} \bibinfo{year}{2020}\natexlab{}.
\newblock \showarticletitle{A comprehensive survey on transfer learning}.
\newblock \bibinfo{journal}{\emph{Proc. IEEE}} \bibinfo{volume}{109}, \bibinfo{number}{1} (\bibinfo{year}{2020}), \bibinfo{pages}{43--76}.
\newblock


\bibitem[Zou et~al\mbox{.}(2022)]%
        {zou2022AESM2}
\bibfield{author}{\bibinfo{person}{Xinyu Zou}, \bibinfo{person}{Zhi Hu}, \bibinfo{person}{Yiming Zhao}, \bibinfo{person}{Xuchu Ding}, \bibinfo{person}{Zhongyi Liu}, \bibinfo{person}{Chenliang Li}, {and} \bibinfo{person}{Aixin Sun}.} \bibinfo{year}{2022}\natexlab{}.
\newblock \showarticletitle{Automatic expert selection for multi-scenario and multi-task search}. In \bibinfo{booktitle}{\emph{Proceedings of the 45th International ACM SIGIR Conference on Research and Development in Information Retrieval}}. \bibinfo{pages}{1535--1544}.
\newblock


\end{thebibliography}


\end{document}